\documentclass[mathpazo]{csiam-am}
\renewcommand\thisnumber{x}
\renewcommand\thisyear {20xx}
\renewcommand\thismonth{xx}
\renewcommand\thisvolume{x}
\usepackage{natbib} 
\usepackage{lipsum}
\usepackage[T1]{fontenc}
\usepackage{microtype}
\usepackage{graphicx}
\usepackage{multirow}
\usepackage{amsmath,amssymb,amsfonts}
\usepackage{amsthm}
\usepackage{mathtools}    
\usepackage{xr-hyper}                 
\usepackage{subfigure}
\usepackage{relsize}
\usepackage{hyperref}      
\usepackage{cleveref}       
\usepackage{wrapfig}
\usepackage{dcolumn}
\usepackage{xcolor} 
\usepackage{geometry}
\usepackage{mathrsfs}
\usepackage[title]{appendix}
\usepackage{xcolor}
\usepackage{textcomp}
\usepackage{manyfoot}
\usepackage{booktabs}
\usepackage{algorithmicx}
\usepackage{algpseudocode}
\usepackage{listings}
\usepackage[most]{tcolorbox}
\usepackage{wrapfig}
\usepackage{booktabs}
\usepackage[linesnumbered,ruled,vlined]{algorithm2e}
\usepackage{pgfplots}
\pgfplotsset{width=0.9\textwidth, height=0.6\textwidth}
\usepackage{tikz}
\usepackage{caption}
\usepackage{newunicodechar}
\newunicodechar{∑}{\ensuremath{\sum}} 
\newunicodechar{≤}{\ensuremath{\le}}
\newunicodechar{≥}{\ensuremath{\ge}}
\newunicodechar{→}{\ensuremath{\to}}
\newunicodechar{−}{-}               
\newunicodechar{ℤ}{\ensuremath{\mathbb{Z}}}
\newunicodechar{ℝ}{\ensuremath{\mathbb{R}}}
\newunicodechar{ℕ}{\ensuremath{\mathbb{N}}}
\newunicodechar{ℚ}{\ensuremath{\mathbb{Q}}}
\newunicodechar{∑}{\ensuremath{\sum}}  
\newunicodechar{Σ}{\ensuremath{\sum}}
\newunicodechar{τ}{\ensuremath{\tau}}
\newunicodechar{ₙ}{\ensuremath{_{n}}}
\newunicodechar{·}{\ensuremath{\cdot}}
\definecolor{keywordcolor}{rgb}{0.7, 0.1, 0.1}   
\definecolor{tacticcolor}{rgb}{0.0, 0.1, 0.3}    
\definecolor{commentcolor}{rgb}{0.4, 0.4, 0.4}   
\definecolor{stringcolor}{rgb}{0.5, 0.3, 0.2}    
\definecolor{symbolcolor}{rgb}{0.1, 0.2, 0.7}    
\definecolor{sortcolor}{rgb}{0.1, 0.5, 0.1}      
\definecolor{attributecolor}{rgb}{0.7, 0.1, 0.1} 
\definecolor{errorcolor}{rgb}{1, 0, 0}           
\newcommand{\lean}[1]{\lstinline[language=lean, mathescape=true]{#1}}

\begin{document}

\title{Translating Informal Proofs into Formal Proofs Using  a Chain of States}


\author[Z. Wang et~al.]{Ziyu Wang\affil{1}\comma,
      Bowen Yang\affil{1},~Chenyi Li\affil{1},~Yuan Zhang\affil{1},~Shihao Zhou\affil{1},~Bin Dong\affil{1},~Zaiwen Wen\affil{1}\corrauth}
\address{\affilnum{1}\ School of Mathematical Sciences,
         Peking University,
         Beijing 100871, P.R. China.}

%
%
\emails{{\tt wangziyu-edu@stu.pku.edu.cn} (Z. Wang), {\tt yangbowen@stu.pku.edu.cn} (B. Yang), {\tt 2200010879@stu.pku.edu.cn} (S. Zhou), {\tt lichenyi@stu.pku.edu.cn} (C. Li), {\tt zy1002@stu.pku.edu.cn} (Y. Zhang), {\tt dongbin@math.pku.edu.cn} (B. Dong), {\tt wenzw@pku.edu.cn} (Z. Wen)}

\begin{abstract}
We address the problem of translating informal mathematical proofs expressed in natural language into formal proofs in Lean4 under a constrained computational budget. Our approach is grounded in two key insights. First, informal proofs tend to proceed via a sequence of logical transitions—often implications or equivalences—without explicitly specifying intermediate results or auxiliary lemmas. In contrast, formal systems such as Lean require an explicit representation of each proof state and the tactics that connect them. Second, each informal reasoning step can be viewed as an abstract transformation between proof states, but identifying the corresponding formal tactics often requires nontrivial domain knowledge and precise control over proof context. To bridge this gap, we propose a two-stage framework. Rather than generating formal tactics directly, we first extract a Chain of States (CoS), a sequence of intermediate formal proof states aligned with the logical structure of the informal argument. We then generate tactics to transition between adjacent states in the CoS, thereby constructing the full formal proof. This intermediate representation significantly reduces the complexity of tactic generation and improves alignment with informal reasoning patterns. We build dedicated datasets and benchmarks for training and evaluation, and introduce an interactive framework to support tactic generation from formal states. Empirical results show that our method substantially outperforms existing baselines, achieving higher proof success rates.
\end{abstract}

\ams{68V15, 68V20
}
\keywords{Automated Theorem Proving, Auto-formalization, Lean 4, Chain of States}

\maketitle

\section{Introduction}
\label{introduction}
In contemporary mathematics, the increasing complexity of formal arguments has made proof verification both labor-intensive and error-prone. Ensuring the correctness of a mathematical result often requires substantial time and meticulous examination by domain experts, posing a challenge for traditional peer review. To address these difficulties, interactive theorem provers such as Lean \citep{de2015lean,moura2021lean} and Isabelle \citep{paulson1994isabelle} have been developed. These systems allow users to express mathematical reasoning in a formal language and obtain fully machine-checked verification. Despite their potential, constructing formal proofs remains highly nontrivial, demanding deep familiarity with both the underlying mathematics and the proof assistant’s syntax and semantics. Meanwhile, large language models (LLMs) such as Deepseek-R1 \citep{deepseekai2025} have demonstrated remarkable reasoning capabilities across a wide range of natural-language and code-related tasks \citep{brown2020language,feng2020codebert,liu2024deepseek}, spurring efforts to integrate them into formal systems such as Lean and Isabelle for automatic theorem proving (ATP) \citep{li2024survey}. 

Automatic theorem proving focuses on generating formal proofs based on a formalized problem statement. Recent approaches commonly cast this task as a Markov decision process, enhancing the reasoning capabilities of provers through tree-based symbolic search \cite{lample2022hypertree}. LLMs are leveraged to predict tactics or local proof steps within interactive theorem proving environments. Leandojo~\citep{Yang2023Leandojo} integrates tactic-based prediction with interactive Lean proof construction. DeepSeek-Prover-V1.5~\citep{deepseekproverv15} combines reinforcement learning with RMaxTS tree search to improve tactic generation. InternLM2.5-StepProver \cite{wu2024internlm2} trains a critic model to perform expert iteration and guide the model to search for deeper proofs. BFS-Prover \cite{xin2025bfs} utilizes the best-first tree search to explore the complex proof space underexplored. In addition to search-based methods, recent research explores end-to-end strategies \citep{xin2024deepseek} that generate full proof trajectories or structured proof trees directly from problem statements. 

Auto-formalization focuses on translating natural language mathematical content into formal statements or proofs within interactive theorem provers. For the task of statement translation, recent research has adopted a range of strategies. One line of work utilizes LLMs with few-shot in-context learning to infer formal statements from informal mathematical descriptions~\citep{wu2022autoformalization,patel2023new,zhou2024don,agrawal2022towards, lu2024formal}. Another approach improves translation accuracy by fine-tuning LLMs on aligned pairs of informal and formal statements~\citep{azerbayev2023proofnet,jiang2023multilingual, liu2025atlas}, thereby reducing reliance on prompt engineering. The LeanWorkBook project~\citep{ying2024leanworkbook} translates a large corpus of natural language mathematical problems into Lean through expert-in-the-loop iterative refinement, resulting in a high-quality dataset. However, when translating informal proofs, there is often a significant mismatch between steps in natural language proofs and tactic steps. The DSP framework~\citep{jiang2022draft} addresses this by leveraging context-aware learning to extract semantically important intermediate steps from LLM-generated informal proofs, which are then completed using automated theorem proving tools such as Isabelle's sledgehammer. On the Lean side, LEGO-Prover~\citep{wang2023lego} decomposes proofs into sub-lemmas, constructs a structured proof database, and applies a retrieval-augmented generation (RAG) approach to guide formal proof synthesis.

\begin{figure}[ht]
  \captionsetup{justification=centering,singlelinecheck=false}
  \centering
  \includegraphics[width=0.8\columnwidth,height=0.45\textwidth]{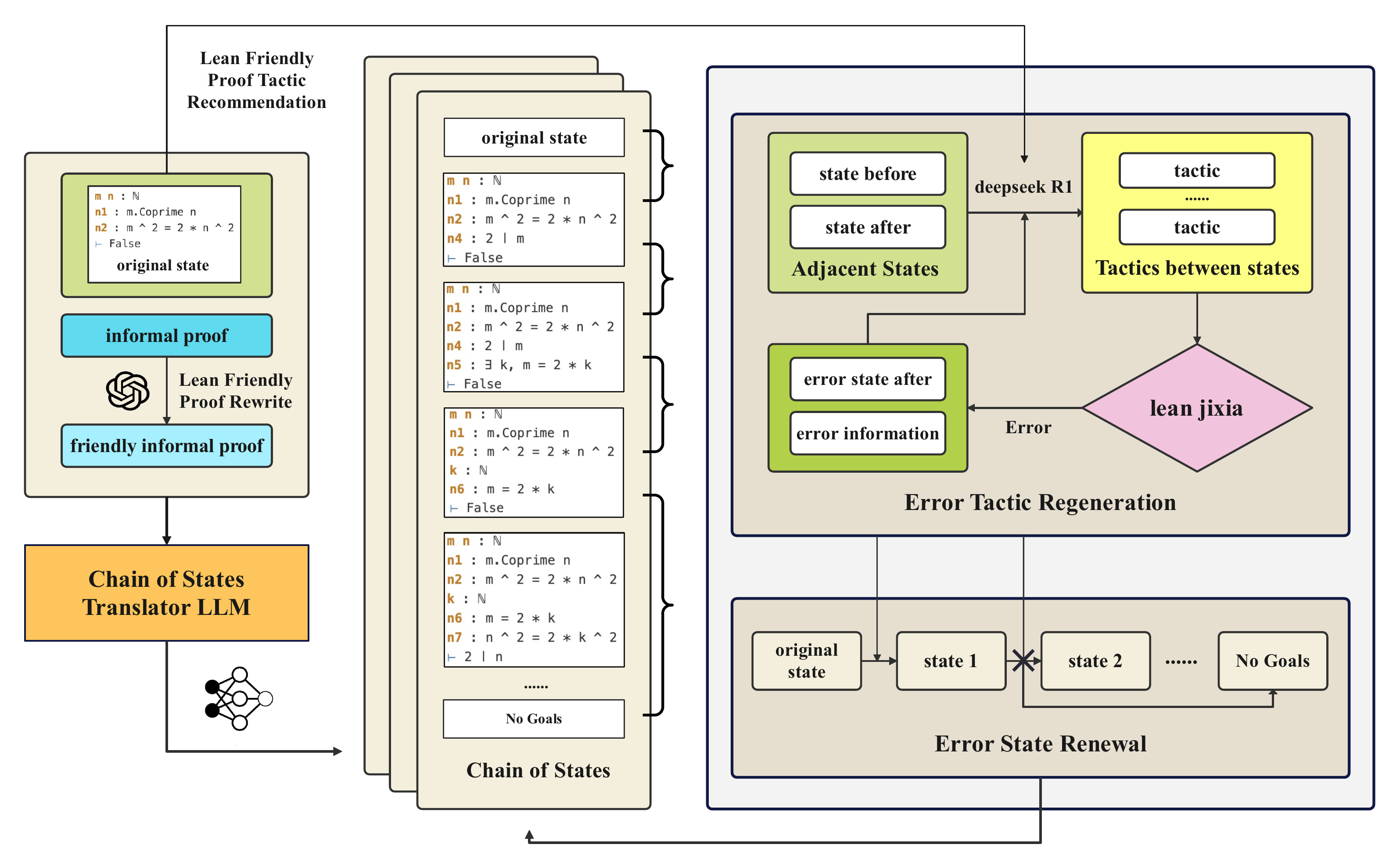}
  \caption{CoSProver}
  \label{fig:example}
\end{figure}

To address the semantic and operational mismatch between informal mathematical proofs and their formal counterparts in systems like Lean, we propose a two-stage translation framework centered on the concept of a Chain of States (CoS). Instead of generating formal tactics directly from informal text, our framework first extracts a semantically aligned sequence of intermediate formal proof states. Tactics are then synthesized between adjacent states to construct complete formal proofs. This intermediate abstraction narrows the reasoning gap, reduces the complexity of tactic prediction, and improves alignment with human reasoning styles. To implement this framework effectively, we introduce robust error correction mechanisms, Lean-friendly rewriting strategies, and data construction pipelines using metaprogramming. Despite requiring significantly fewer prover invocations than existing methods, our approach achieves state-of-the-art performance on challenging benchmarks such as MiniF2F—demonstrating both high efficiency and strong generalizability. This makes our system especially suitable for real-world scenarios with limited computational resources. An overview of our method is shown in Figure \ref{fig:example}. Our contributions are as follows.\\
1. \textbf{Informal proof - CoS semantic alignment framework.} We propose a novel intermediate abstraction, Chain of States, which semantically aligns informal proof reasoning with formal proof structures. This enables more accurate tactic generation, reduces logical ambiguity, and improves interpretability and proof modularity.\\
2. \textbf{Error-aware tactic generation with Lean-friendly rewriting.} We develop a hybrid tactic generation pipeline that combines LLM-based generation with symbolic provers, enhanced by modules for error-aware regeneration and state renewal. Furthermore, we incorporate Lean-specific heuristics by rewriting informal proofs into Lean-friendly forms, improving both proof success rate and efficiency.\\
3. \textbf{Dataset and benchmark construction via metaprogramming.} We construct a large-scale dataset of informal–formal proof pairs by augmenting term-based and tactic-based proofs from Mathlib and LeanWorkbook using metaprogramming on elaboration trees. We also establish comprehensive benchmarks across multiple mathematical domains, enabling rigorous evaluation of semantic alignment and proof validity.

\section{Introduction to Formal Languages and Lean}

\subsection{Overview of Formal Languages}

Formal proof assistants are interactive systems designed to construct, verify, and manage rigorous mathematical proofs using precise logical foundations. They allow users to formalize definitions, theorems, and proofs in a machine-checkable way, ensuring correctness through small trusted kernels and type-theoretic or logical frameworks. Over the past decades, several prominent systems have emerged, each with its own strengths and foundational logic. \textbf{Coq}, based on the calculus of inductive constructions, excels in constructive mathematics and program verification. \textbf{Isabelle/HOL} adopts classical higher-order logic and offers powerful automation via its Isar proof language. \textbf{HOL Light}, known for its minimalist trusted core, has been used in highly rigorous mathematical formalizations. \textbf{Agda} emphasizes dependent types and interactive proof development through functional programming constructs. 

\textbf{Lean 4} is a modern interactive theorem prover based on a dependently typed logical framework. Compared to earlier systems, Lean 4 is developed not only for formal proof construction but also with support for general-purpose programming and metaprogramming. A distinguishing feature of Lean is its rapidly expanding formalization community, which maintains the open-source library \textit{mathlib}. Mathlib is one of the most extensive and actively maintained collections of formalized mathematics, covering core areas such as algebra, analysis, topology, number theory, and linear algebra. It is developed collaboratively by a large number of contributors and adopts a modular organization. The library supports both elementary topics and advanced research-level formalizations.

In recent years, large language models (LLMs) have been increasingly applied to formal reasoning tasks, using interactive theorem provers as target systems. These applications include tactic prediction, proof state generation, and the translation of informal mathematical texts into formal proofs. Formal systems provide a structured environment and verifiable logic foundation, which are beneficial for training and evaluating LLMs in mathematical reasoning. Among existing proof assistants, Lean 4 has become one of the most commonly used platforms in LLM-related research, due to the availability of a large formal corpus and an active community that supports integration with AI-based methods.

\subsection{State and Tactic in Lean}
In Lean, the concepts of state and tactic play a fundamental role in the interactive proof process. We introduce the core ideas behind these concepts.

\textbf{State} is a structure representing the current proof environment, consisting of several \textbf{goals}. Each goal corresponds to a conclusion needing proof. For a goal $\mathcal{G}$ with target conclusion $g$ and hypotheses $h_1, h_2, \dots, h_m$, it is written as ${h_1, h_2, \dots, h_m} \vdash g$. A state has two representations, as a \textbf{formal statement} in Lean code that defines the problem's hypotheses and target conclusions; second, as an \textbf{infoview state} in Lean 4's infoview that displays conclusions and conditions at the current cursor position. Mathematically, a state $\mathcal{S}$ is $\mathcal{S} = \{\mathcal{G}_1, \mathcal{G}_2, \dots, \mathcal{G}_n\}$, where each $\mathcal{G}_k$ has the form $\{h_{k1}, h_{k2}, \dots, h_{km_k}\} \vdash g_k$, with $\{h_{k1}, h_{k2}, \dots, h_{km_k}\}$ being hypotheses and $g_k$ the target conclusion.

\textbf{Tactic} is a rule or operation that transforms the current state into a new one by modifying the goals. A tactic $\mathcal{T}$ can be seen as a function mapping state $\mathcal{S}$ to a new state $\mathcal{S}'$. Each tactic $\mathcal{T}$ addresses specific logical structures or applies particular theorems. In Lean, theorem proofs can be organized in tactic form (\textit{tactic-based proofs}). To prove a statement $\mathcal{S}_0$, the proof consists of a sequence of tactics $\{\mathcal{T}_1, \mathcal{T}_2, \dots, \mathcal{T}_n\}$ applied in order. After the last tactic, if the resulting state has no remaining goals, the proof is finished. The whole proof can be represented mathematically as:
\[
\mathcal{S}_0 \xrightarrow{\mathcal{T}_1} \mathcal{S}_1\xrightarrow{\mathcal{T}_2} \dots \xrightarrow{\mathcal{T}_n} \mathcal{S}_n
\]

When a proof is organized in tactic form, it is represented as a sequence of tactics $\mathcal{P}^F=\{\mathcal{T}_1, \mathcal{T}_2, \dots, \mathcal{T}_n\}$. The interplay between state and tactic forms the core of Lean's interactive proof process. When writing the formal proofs, human will continuously observe and understand the current state while searching for appropriate tactics to modify it.

\subsection{Writing Formal Proofs in Lean}

In Lean, there are typically two ways to write a proof for a problem: \textbf{term-based proofs} and \textbf{tactic-based proofs}. A term-based proof directly constructs the target goal using nested expressions that represent the proof object. This style is prevalent in \textbf{mathlib}, where it is especially suitable for proving corollaries or applying existing theorems in a concise manner. Term-based proofs offer composability and close correspondence with constructive type theory, but they can be difficult to manipulate interactively, particularly for longer or more complex arguments.

Tactic-based proofs, in contrast, construct the proof incrementally by applying tactics that transform the current goal state. This style is more commonly used in interactive sessions by both human users and large language models, as it provides a more transparent, stepwise process. Each tactic corresponds to a logical transformation and results in a new proof state. This sequential structure makes it easier to follow the reasoning process and to manage proof development interactively.

For human users writing tactic-based proofs, there are typically two main organizational strategies. The first is the \textbf{step-by-step} approach, where the user progressively and carefully simplifies the current goal through a linear sequence of tactic applications. This is effective for relatively simple, straightforward problems with direct proof paths and allows the user to focus on small logical steps. The second strategy is known as the \textbf{Blueprint} method, which is more suitable for complex theorems. In this approach, users first construct a clear high-level skeleton of the proof—often involving modular auxiliary lemmas or subgoals—and then complete each part independently. This helps maintain global logical structure and supports collaborative formalization, as different users can work on different parts of the blueprint concurrently.

In the context of large language models for formal theorem proving, the two human-centered strategies correspond to two distinct problem paradigms: \textbf{Automatic Theorem Proving} (ATP) and \textbf{Auto-formalization}. ATP focuses on tactic prediction based on the current formal state. Formally, given a state $\mathcal{S}_p$, the goal is to generate one or more tactics $\{\mathcal{T}_1, \dots, \mathcal{T}_k\}$ that lead to a successor state $\mathcal{S}_n$:
\[
\mathcal{S}_p \rightarrow \mathcal{T}_{p+1} \quad \text{or}
\quad \mathcal{S}_p \rightarrow \{\mathcal{T}_{p+1},\mathcal{T}_{p+2},\cdots \mathcal{T}_{n}\}
\]
This task can be modeled as a Markov Decision Process (MDP), where proof states \(\mathcal{S}_p\) are states in the MDP, tactics are actions, and the goal is to find a path that eventually leads to a terminal state with no remaining goals. The MDP formulation enables the use of search algorithms, reinforcement learning, and trajectory-based data to guide tactic synthesis. Structurally, ATP aligns well with the step-by-step method used by humans.

Auto-formalization, in contrast, aims to translate informal proofs into complete formal tactic sequences. Given an initial formal state $\mathcal{S}_0$ and an informal proof $\mathcal{P}^I$, the objective is to generate a complete tactic-based proof $\mathcal{P}^F = \{\mathcal{T}_1, \mathcal{T}_2, \dots, \mathcal{T}_n\}$. This task requires aligning informal reasoning with formal proof structures, making it relatively more challenging compared to Automatic Theorem Proving. 

\section{Motivation}
\label{Motivation}

\subsection{Differences between Informal and Formal Proofs}
\label{sec: differences}
The primary challenge of the translation task arises from the substantial gaps between the informal proofs and formal proofs. In this subsection, we analyze these issues from the perspectives of semantic interpretation and proof strategy. From a semantic interpretation perspective, we can see that informal proofs directly reflect the change of different states, while the formal proofs focus on giving the transition reasons between different states. Moreover, informal and formal proofs favor different techniques, reflecting the differing burdens on human and machine provers.

We begin by understanding proofs semantically. Informal proofs typically deduce only the key intermediate conclusions, rather than detailing every minor result. Most steps take the form of a simple implication, phrased as "assuming A, we conclude B" or "to complete the argument, it suffices to prove C under D". By chaining these intermediate conclusions, we see exactly how each main step leads to the next, keeping the proof clear and easy to follow. 
Another key characteristic of informal proofs is the omission of minor implication steps, as they are typically considered self-evident and do not impede the reader’s comprehension.
However, formal proofs are tactic-centered, requiring rigorous justification for each state transition through explicitly written tactics. The intermediate proof states themselves are implicit and can only be accessed through auxiliary tools such as the "infoview" in interactive theorem provers.

\begin{figure}[ht]
    \centering
    \includegraphics[width=0.9\linewidth, height=0.35\textwidth]{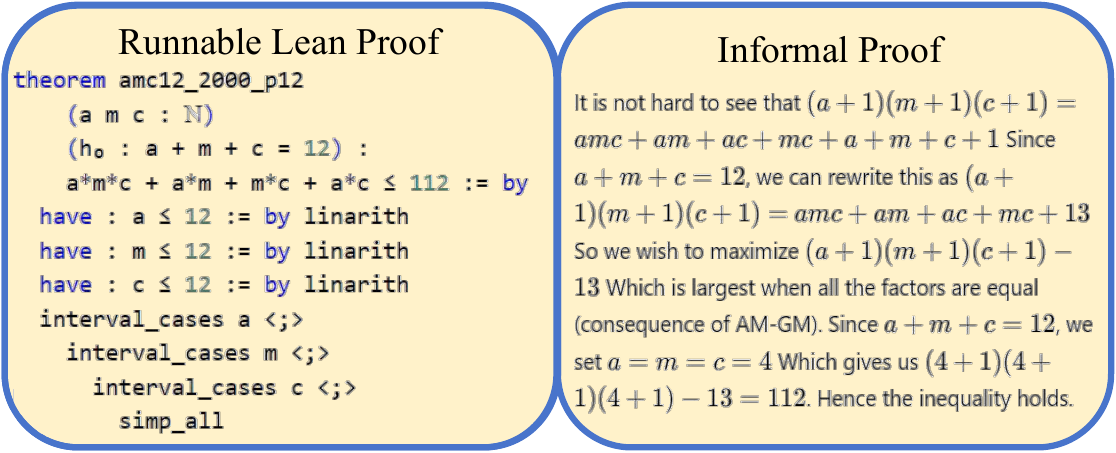}
    \caption{Differences between informal proof and formal proof.}
    \label{fig:difference}
\end{figure}

Informal proofs also differ significantly from formal ones in their choice of proof strategies. For instance, proof by exhaustion, which involves splitting assumptions into numerous cases and verifying each individually, is commonly employed in formal systems, as machines can process cases in parallel and generate error-free deductions. For human provers, however, this method is tedious and prone to oversight, particularly with corner cases. An example is presented in Figure \ref{fig:difference}. Conversely, strategies such as "without loss of generality" are well suited for human reasoning but resist formalization. Moreover, while humans often rely on heuristic simplification strategies to prove equations, formal systems like Lean can validate them directly via symbolic computation, avoiding transformations.

A few types of these differences are listed in Table \ref{tab:differences}. More examples are given in Section \ref{examples}. These differences impose challenges in translating informal proofs into formal proofs. Based on these two observations, we propose the concept of \textbf{Lean-friendly natural language proof}. These informal proofs are closer in proving methodology to formal proofs, as they contain a more detailed deduction process and adopt Lean-friendly proving techniques.

\begin{table}[ht]
    \centering
    \caption{Examples of differences between informal proof and formal proof in MiniF2F.}
    \label{tab:differences}
    \begin{tabular}{l c}
    \toprule
    Difference type & Occurred cases \\ 
    \midrule
    Proof by exhaustion  & Number Theory (related to $\mathbb{N}$) \\ 
    Proof by calculation & Divisors Calculation\\
    Type Transformation  & Algebra (related to $\mathbb{R}$ and $\mathbb{N}$)\\ 
    Logical Leaps & Using secondary conclusions\\ 
    \bottomrule
    \end{tabular}
\end{table}

\subsection{Lean-Friendly Informal Proof Rewrite}
\label{leanfriend}
Motivated by the recurring differences summarized in Table~\ref{tab:differences}, we perform a lightweight preprocessing of both the informal proofs and the tactic-generation prompts, aligned to the formal statement. This step precedes any downstream search or sequencing mechanism and aims to reduce translation brittleness by surfacing tactic-relevant structure without altering the mathematical content.

All natural language proofs undergo preprocessing with four key operations. First, we remove redundant explanatory text lacking logical content. Second, we align with the formal statement’s mathematical modeling: since MiniF2F problems often involve application scenarios where the formal statements already incorporate modeling, we eliminate duplicate modeling descriptions. Third, for entirely non-rigorous proofs, we regenerate the natural language version. Finally, we replace ambiguous terms (e.g., “similarly,” “obviously”) with explicit reasoning steps and clarify proof structures (e.g., contradiction, induction) with their precise conditions and conclusions.

Furthermore, some problems admit simpler proofs in Lean than their natural language counterparts. When exploring multiple candidate proof attempts, we retain the simplest script that closes the goal immediately after the initial state. For such cases, we adjust the retry prompts to suggest Lean-specific methods. For instance, in secondary school number theory problems where natural language proofs fix variable ranges before discussing properties, Lean often prefers exhaustive enumeration within those ranges. Our prompt advises ignoring the original proof for such range-constrained problems and instead performing case enumeration, effectively aligning with Lean’s optimization without altering the informal proof’s content~\ref{exhaustion}.

\subsection{Examples of Informal-Formal Proof Differences}
\label{examples}
We identify four typical distinctions between informal and formal proofs—proof by exhaustion, proof by calculation, type transformation, and logical leaps—and illustrate each with an example.

\subsubsection{Proof by exhaustion}
\label{exhaustion}
\begin{figure}[ht]
  \centering
  \begin{tcolorbox}[
    colback=gray!10,    
    colframe=black,     
    boxrule=0.5pt,      
    arc=1mm,            
    left=2mm, right=2mm,
    top=0.2mm, bottom=0.2mm
  ]
  \begin{lstlisting}[language=Lean]
theorem amc12_2000_p12
    (a m c : ℕ)
    (h₀ : a + m + c = 12) :
    a*m*c + a*m + m*c + a*c ≤ 112 := by
    have : a ≤ 12 := by linarith
    have : m ≤ 12 := by linarith
    have : c ≤ 12 := by linarith
    interval_cases a <;>
      interval_cases m <;>
        interval_cases c <;>
          simp_all
  \end{lstlisting}
  \end{tcolorbox}
  \begin{tcolorbox}[
    colback=gray!10,
    colframe=black,
    boxrule=0.5pt,
    arc=1mm,
    left=2mm, right=2mm,
    top=2mm, bottom=2mm
  ]
  First note that
  \[
    0 \le a,m,c \le 12,
    \quad
    a + m + c = 12.
  \]
  We then perform a nested split on \(a\in\{0,\dots,12\}\),  
  then on \(m\in\{0,\dots,12\}\),  
  and finally on \(c\in\{0,\dots,12\}\).  
  In each of the finitely many subcases, a direct arithmetic simplification shows  
  \[
    a\,m\,c + a\,m + m\,c + a\,c \;\le\; 112.
  \]
  Hence the inequality holds in all cases.  
  \end{tcolorbox}
  \caption{Formal proof of \texttt{amc12\_2000\_p12} and its lean-friendly informal proof.}
  \label{fig:amc12-proof}
\end{figure}

As explained in Section 2, proof by exhaustion is a technique which formalization method often use to deal with problems related to natural numbers. We explain the example in detail here. The formal proof and its corresponding lean-friendly informal proof are in Figure \ref{fig:amc12-proof}. It discusses all the possible cases which $a$, $m$,$c$ can take and proof the corresponding inequality. The tactic \lean{interval_cases a <;>} means to apply the tactic after \lean{<;>} on all cases. This search area is small for machine but huge for human.

\begin{figure}[ht]
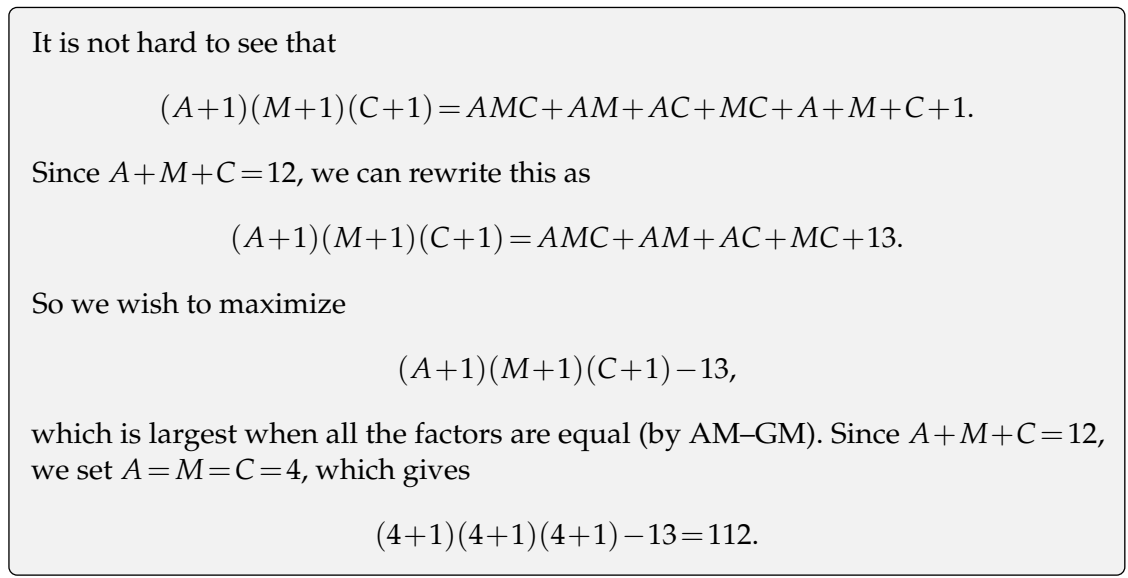

  \centering
  \begin{tcolorbox}[
    colback=gray!10,
    colframe=black,
    boxrule=0.5pt,
    arc=1mm,
    left=2mm, right=2mm,
    top=2mm, bottom=2mm
  ]
  It is not hard to see that 
  \[
    (A+1)(M+1)(C+1)
    = AMC + AM + AC + MC + A + M + C + 1.
  \]
  
  Since \(A+M+C=12\), we can rewrite this as
  \[
    (A+1)(M+1)(C+1)
    = AMC + AM + AC + MC + 13.
  \]
  
  So we wish to maximize
  \[
    (A+1)(M+1)(C+1) - 13,
  \]
  which is largest when all the factors are equal (by AM–GM).  
  Since \(A+M+C=12\), we set \(A=M=C=4\), which gives
  \[
    (4+1)(4+1)(4+1) - 13 = 112.
  \]
  \end{tcolorbox}
  \caption{Lean-unfriendly informal proof of \texttt{amc12\_2000\_p12}}
  \label{fig:amc12_informal_proof}
\end{figure}

The natural language proof is based on a core equality $(a+1)(m+1)(c+1) = amc+am+ac+mc+a+m+c+1$ and performs secondary conclusion AM-GM on the target inequality as Figure \ref{fig:amc12_informal_proof}. 
 Hence, we can see that for certain problems, machine assisted proving can help human do lots of tedious calculation work on discussing the cases at the same time.

\subsubsection{Proof by calculation}
This situation typically arises when humans cannot directly evaluate a summation to concrete numerical values, whereas machines can. An example is provided in Figure \eqref{fig:mathd_numbertheory_427}.
Lean utilize "rfl" to perform calculation by the definition of \textit{Nat.divisors} in this case as Figure \ref{fig:mathd_numbertheory_427}, which is quite straightforward and simple in formalization. However, the natural language proof is somewhat more tedious to deal with the budget of calculation.
The human friendly informal proof compute $a$ by applying the well-known result that the sum of all a number’s divisors can be expressed directly from its prime factorization, which is given in Figure \ref{fig:mathd_numbertheory_427_informal}. Formalizing this result in a proof assistant, however, involves additional complexity.

\begin{figure}[ht]
  \centering
  \begin{tcolorbox}[
    colback=gray!10,
    colframe=black,
    boxrule=0.5pt,
    arc=1mm,
    left=2mm, right=2mm,
    top=0.2mm, bottom=0.2mm
  ]
  \begin{lstlisting}[language=Lean,
  mathescape=true,
  columns=fullflexible, keepspaces=true, showstringspaces=false,
  literate={∑}{{$\sum$}}1 {Σ}{{$\sum$}}1 {ℕ}{{$\mathbb{N}$}}1
           {₀}{{\(_{0}\)}}1 {₁}{{\(_{1}\)}}1 {₂}{{\(_{2}\)}}1 {₃}{{\(_{3}\)}}1 {₄}{{\(_{4}\)}}1
           {₅}{{\(_{5}\)}}1 {₆}{{\(_{6}\)}}1 {₇}{{\(_{7}\)}}1 {₈}{{\(_{8}\)}}1 {₉}{{\(_{9}\)}}1]
theorem mathd_numbertheory_427 (a : ℕ)
    (h₀ : a = (∑ k in (Nat.divisors 500), k)) :
    ∑ k in Finset.filter (fun x => Nat.Prime x) (Nat.divisors a), k = 25 := by
      rw [h₀]
      rfl
\end{lstlisting}
\end{tcolorbox}
\begin{tcolorbox}[
    colback=gray!10,
    colframe=black,
    boxrule=0.5pt,
    arc=1mm,
    left=2mm, right=2mm,
    top=2mm, bottom=2mm
  ]
  We substitute the value of $a$ into the target equation. A straightforward calculation, using the definition of divisors, produces the desired result.
  \end{tcolorbox}
  
  \caption{Formal proof and lean-friendly informal proof of mathd\_numbertheory\_427 }
  \label{fig:mathd_numbertheory_427}
\end{figure}

\begin{figure}[ht]
  \centering
  \begin{tcolorbox}[
    colback=gray!10,    
    colframe=black,     
    boxrule=0.5pt,      
    arc=1mm,            
    left=2mm, right=2mm,
    top=2mm, bottom=2mm
  ]
  First, we find \(A\). The prime factorization of \(500\) is \(2^2 \cdot 5^3\). Therefore,
  \[
    A \;=\; (1 + 2 + 2^2)\,(1 + 5 + 5^2 + 5^3)
    \;=\; 7 \times 156.
  \]
  To see why \((1+2+2^2)(1+5+5^2+5^3)\) equals the sum of the divisors of \(500\),
  note that expanding it yields exactly all \(2^2\cdot5^3\) divisors once.

  Next, factorize \(7 \times 156 = 7 \times 2^2 \times 3 \times 13\).
  The sum of the prime divisors of \(A\) is
  \[
    2 + 3 + 7 + 13 = 25.
  \]
  \end{tcolorbox}
  \caption{Lean‐unfriendly informal proof of mathd\_numbertheory\_427}
  \label{fig:mathd_numbertheory_427_informal}
\end{figure}

\subsubsection{Type transformation}
In Lean formalization, each variable and numeral is assigned a fixed type. Consequently, to reinterpret a term in another type, one must apply explicit type‐casting methods. For example, the expression
\(
(1/3) : \mathbb{N}
\)
evaluates to
\(
0 : \mathbb{N},
\)
whereas
\(
(1/3) : \mathbb{R}
\)
yields the real fraction
\(
\frac{1}{3} : \mathbb{R}.
\)
In lean, numbers $1, 2, \cdots$ are taken as natural numbers $\mathbb{N}$ by default. Therefore, formal proofs require additional steps to manage these type transformations. A simple yet representative case is given below in Figure \ref{fig:mathd_algebra_114}. Since the one-third power needs to deal with the power between natural numbers and real numbers, the formal proof and its corresponding informal proof show that detailed process to handle the one-third power part.
\begin{figure}[ht]
    \centering
    \begin{tcolorbox}[
  colback=gray!10,    
  colframe=black,     
  boxrule=0.5pt,      
  arc=1mm,            
  left=2mm, right=2mm,
  top=0.2mm, bottom=0.2mm
]\begin{lstlisting}[language=Lean]
theorem mathd_algebra_114 (a : ℝ) (h₀ : a = 8) :
    (16 * (a^2)^((1 : ℝ) / 3))^((1 : ℝ) / 3) = 4 := by
  rw [h₀]
  norm_num; ring_nf
  have : (64 : ℝ) ^ ((1 : ℝ) / 3) = 4 := by
    rw [show (64 : ℝ) = 4^3 by norm_num]
    rw [← rpow_nat_cast, ← rpow_mul (by norm_num)]
    norm_num
  rw [this]
  simp; norm_num
  rw [this]
\end{lstlisting}
\end{tcolorbox}
\begin{tcolorbox}[
  colback=gray!10,    
  colframe=black,     
  boxrule=0.5pt,      
  arc=1mm,            
  left=2mm, right=2mm,
  top=0.2mm, bottom=2mm
]
We wish to show \(
    \bigl(16\,(a^2)^{\tfrac13}\bigr)^{\tfrac13} = 4
  \)\text{ under the assumption} $a = 8$. We substitute the value of $a$ to get
    \(
      \bigl(16\,(a^2)^{\tfrac13}\bigr)^{\tfrac13}
      = \bigl(16\,(8^2)^{\tfrac13}\bigr)^{\tfrac13}.
    \)
   Since \(8^2 = 64\), this becomes
    \(
      \bigl(16 \cdot 64^{\tfrac13}\bigr)^{\tfrac13}.
    \)
    Note that \(64 = 4^3\), we have
    \(
      64^{\tfrac13}
      = (4^3)^{\tfrac13}
      = 4^{\,3\cdot\tfrac13}
      = 4,
    \)
    so the expression simplifies to
    \(
      64^{\tfrac13} = 4,
    \)
   which is the desired result.
\end{tcolorbox}
    \caption{Formal proof of mathd\_algebra\_114 and its corresponding informal proof.}
    \label{fig:mathd_algebra_114}
\end{figure}

\begin{figure}[ht]
    \centering
\begin{tcolorbox}[
  colback=gray!10,    
  colframe=black,     
  boxrule=0.5pt,      
  arc=1mm,            
  left=2mm, right=2mm,
  top=0.2mm, bottom=0.2mm
]
Note that $a^2 = 64$ and $\sqrt[3]{64} = 4$. Therefore, $\left(16\sqrt[3]{a^2}\right)^{\frac {1}{3}} = \left(16 \times 4\right)^{\frac{1}{3}} = 64^\frac{1}{3} = 4.$
\end{tcolorbox}
    \caption{Lean‐unfriendly informal proof of  mathd\_algebra\_114.}
    \label{fig:mathd_algebra_114_informal}
\end{figure}

However, the informal proof is quite straightforward as Figure \ref{fig:mathd_algebra_114_informal}. We see that the informal proof does not emphasize the type of these numbers. Most readers easily understand the one-third power and assume the correct numeric types without explicit mention. However, in formal proof, we need to use premises such as \lean{rpow_nat_cast} to deal with the transformation between $\mathbb{N}$ and $\mathbb{R}$. Besides, while $\sqrt[3]{64} = 4$ is explicit in informal proof. Formal proof uses quite a few tactics to give this proof in the \lean{have} part.

For each of the types of problems mentioned above, we outlined the characteristics and provided examples of lean-friendly informal proofs. We used GPT-4o to first classify the problems and then refine the informal proofs accordingly. Furthermore, when the original CoSProver attempted to prove the simplest CoS instances, we incorporated the characteristics of the corresponding type of lean-friendly formal proofs into the prompts for those problems. This encouraged the model to employ proof techniques such as exhaustion and calculation, while also highlighting potential issues related to type conversions.

\subsubsection{Logical leaps}
Some steps in informal proofs are simple and acceptable for human readers, as they can intuitively fill in the omitted parts of the proof. However, these logical leaps often present significant challenges for formal provers, as bridging such gaps is not straightforward and is problem- and domain-specific. In this subsection, we illustrate two relatively complex examples, shown in Figure \ref{fig:mathd_algebra_320_informal} and Figure \ref{fig:imo_1963_p5}, to highlight these difficulties.

\begin{figure}[h]
    \centering
    \begin{tcolorbox}[
  colback=gray!10,    
  colframe=black,     
  boxrule=0.5pt,      
  arc=1mm,            
  left=2mm, right=2mm,
  top=2mm, bottom=2mm
]
Problem: Let \(x\) be a real number and \(a,b,c\) be positive integers such that \(x\ge0\).  Suppose further that
\(
2x^2 = 4x + 9,
x = \frac{a + \sqrt{b}}{c},
c = 2.
\)
Prove that
\(
a + b + c = 26.
\)

Informal Proof: First, we move all terms to one side to get $2x^2 - 4x - 9 = 0.$ Seeing that factoring will not work, we apply the Quadratic Formula: \begin{align*}
x &= \frac{-(-4) \pm \sqrt{(-4)^2 - 4(2)(-9)}}{2 (2)}
= \frac{4 \pm \sqrt{16 + 72}}{4}  = \frac{2 \pm \sqrt{22}}{2}.
\end{align*}Since $x$ is positive, $x$ can be written as $\dfrac{2 + \sqrt{22}}{2},$ so our answer is $2 + 22 + 2 = 26.$
\end{tcolorbox}
    \caption{Lean-unfriendly informal proof of mathd\_algebra\_320}
    \label{fig:mathd_algebra_320_informal}
\end{figure}

In the first step of the informal proof in Figure \ref{fig:mathd_algebra_320_informal}, the value of \(x\) is deduced using the quadratic formula, a lemma not well formalized in the Mathlib4 library. Since this step may not be completed in a single tactic in formal language, it represents a logical leap. Additionally, the conclusion relies on the fact that \(a\) and \(b\) are natural numbers satisfying the given equation. While it is straightforward for humans to see that \(a + \sqrt{b} = 2 + \sqrt{22}\) implies \(a = 2\) and \(b = 22\), this step must be handled using a proof by contradiction in formal language. The only way to confirm this is by testing all possible natural number values and showing that \(a = 2\) is the only reasonable solution. 

\begin{figure}[h]
    \centering
    \begin{tcolorbox}[
  colback=gray!10,    
  colframe=black,     
  boxrule=0.5pt,      
  arc=1mm,            
  left=2mm, right=2mm,
  top=2mm, bottom=2mm
]
Problem: 
Prove that
\(
\cos{\frac{\pi}{7}}-\cos{\frac{2\pi}{7}}+\cos{\frac{3\pi}{7}} = \frac{1}{2}.
\)

Informal Proof: 
Let $\cos{\frac{\pi}{7}}-\cos{\frac{2\pi}{7}}+\cos{\frac{3\pi}{7}}=S$. We have
$$S=\cos{\frac{\pi}{7}}-\cos{\frac{2\pi}{7}}+\cos{\frac{3\pi}{7}}=\cos{\frac{\pi}{7}}+\cos{\frac{3\pi}{7}}+\cos{\frac{5\pi}{7}}.$$

Then, by product-sum formulae, we have

$$S * 2* \sin{\frac{\pi}{7}} = \sin{\frac{2\pi}{7}}+\sin{\frac{4\pi}{7}}-\sin{\frac{2\pi}{7}}+\sin{\frac{6\pi}{7}}-\sin{\frac{4\pi}{7}}=\sin{\frac{6\pi}{7}}=\sin{\frac{\pi}{7}}.$$
Thus $S = 1/2$
\end{tcolorbox}
    \caption{Lean-unfriendly informal proof of imo\_1963\_p5}
    \label{fig:imo_1963_p5}
\end{figure}

For the given proof in Figure \ref{fig:imo_1963_p5}, we are tasked with proving that
\(
\cos{\frac{\pi}{7}} - \cos{\frac{2\pi}{7}} + \cos{\frac{3\pi}{7}} = \frac{1}{2}.
\)
After simplifying the expression for $S$, human provers may use trigonometric identities and symmetry to relate the sum of cosines to a simpler form. The critical step involves using the product-to-sum formula to transform the expression. However, this introduces a logical leap, as the transformation is not immediately intuitive in formal language. The proof relies on recognizing that 
\(
\cos{\frac{\pi}{7}} + \cos{\frac{3\pi}{7}} + \cos{\frac{5\pi}{7}} 
\)
can be expressed in terms of sine functions, which are manipulated using the sum-to-product identities. This omits many detailed calculation steps. The fact that the resulting sine terms simplify to $\sin{\frac{\pi}{7}}$ is crucial. While this step is evident to humans familiar with trigonometric identities, formalizing it requires a detailed series of tactic-based manipulations. The final deduction, $S = \frac{1}{2}$, comes from the fact that $\sin \frac{\pi}{7} \neq 0$, which is another leap in the informal proof. These leaps make it difficult to translate this informal proof into formal.

\section{Generation of Chain of States from Informal Proof}
\label{sec: generation}

\subsection{Aligning the Chain of States with Informal Proofs}
\label{aligning}
 Both informal steps and formal tactics essentially represent mappings between the previous state $\mathcal{S}_p$ and the next state $\mathcal{S}_n$, i.e., $(\mathcal{S}_{p} \rightarrow \mathcal{S}_{n}) \leftrightarrow \mathcal{T}$. 
In natural language proofs, it is important to note that states are not explicitly given. The informal proofs tend to highlight the assumptions that are introduced or how the goal evolves. A complete state is typically stated only at the beginning of a proof or when a lemma is presented independently. However, this suggests that informal proofs and formal proof states are not fundamentally different: at their core, informal proofs still describe transitions between adjacent states, and humans can readily infer the full state at each step, including relevant assumptions and current goals.

In the process of aligning CoS with informal proofs, GPT-4o is utilized to generate informal proofs for each formal proof as input. At the same time, we use GPT-4o to translate each state into natural language, creating a dataset for formal and informal state alignment. Finally, to clarify the differences described in natural language proofs, informal explanations of state changes are generated for each adjacent state, which aligns informal states with informal proofs. The prompts
are given in Appendix. To construct datasets of informal proofs and CoS, complete formal proofs are collected from Leanworkbook \cite{ying2024leanworkbook}, Leanworkbook-plus \cite{goedelprover}, and Mathlib. The CoS and tactics are extracted for each problem. During data cleaning, we find that many theorems in Mathlib consist of only one tactic, leading to generating too few states in the model. As a result, we filter the problems that have two or more states. Additionally, Mathlib is filtered to focus only on theorems relevant to mathematics, as a significant amount of its code pertains to Lean's underlying framework rather than the construction of mathematical theorems, retaining approximately half of the theorems.

Based on the aligned data, we perform supervised fine-tuning to train a model translating informal proof to CoS. Given that Deepseek-R1 is one of the few general models with some capabilities in Lean4, we use Qwen 7B distilled from Deepseek-R1 as the pretrained model to fine-tune the CoS translation data.

\subsection{Data Augmentation: Leveraging Intermediate Tactics to Improve the Chain of States}
\label{data_aug}
To align informal proofs with formal proofs, we first extract chains of proof states from Lean formal proofs. Typically, formal proofs fall into two categories: \textbf{tactic-based proofs} (following a \texttt{"by"} keyword) and \textbf{term-based proofs}. Tactic-based proofs are more common and allow for relatively straightforward extraction of state chains. However, certain tactics—such as \texttt{"have"}—may encapsulate subproofs whose internal state transitions are not directly accessible.
In contrast, term-based proofs are expressed as composite proof terms and are thus inherently indivisible; they do not yield extractable chains of states.
To address these limitations, we design data augmentation techniques that expand internal tactic patterns in traditional proofs—for example, by extracting multiple steps from \texttt{have} constructs—and decompose complex term compositions to expose finer-grained state transitions. This approach enables the inclusion of term-based data that traditional methods overlook, significantly enhancing the coverage and granularity of formal proof-state datasets.

Mathlib proofs often use a term-based style rather than tactic-based proofs. Even when tactics are used, they may combine with complex term compositions or theorem applications, causing logical jumps between adjacent states. This poses challenges for extracting coherent state chains, creating a deviation from typical formal proofs and informal proofs. We propose transforming Mathlib theorems, particularly term-based proofs, into formal proofs following normal logical inference. 
Currently, there is no general method for breaking down term-based proofs or complex theorem compositions. Previous attempts to increase data quantity through "rw" (rewrite) tactic decomposition only handle single tactics, leaving common term compositions unaddressed \citep{nesterov2024leantrace}.
Fortunately, 
extracting states from terms is simpler than decomposing complex terms into equivalent tactics. In Lean, both term-style and tactic-style proofs appear in the elaboration tree, a tree-like structure representing theorem compositions. 

In Lean’s compilation pipeline, the \emph{Elaboration Tree} is the definitive artifact recording every inference and refinement step that transforms high-level surface syntax into kernel-level core terms. Each node \(N\) in this directed acyclic graph carries the tuple
\[
  N = (\mathit{expr},\;\Gamma,\;\Sigma,\;\Delta,\;\mathit{ctx},\;\mathit{res}),
\]
where \(\mathit{expr}\) denotes the intermediate expression node, \(\Gamma\) the local hypothesis context, \(\Sigma\) the global environment of declarations, \(\Delta\) the current metavariable assignment, \(\mathit{ctx}\) the type-checking environment (including unification constraints and type-class resolution state), and \(\mathit{res}\) the fully elaborated core term produced at that node.

During elaboration, Lean systematically resolves omitted information: implicit arguments \(?_i\) are instantiated to \(v_i : \tau_i\), type-class instances of type \(\tau\) are inferred to \(\mathit{inst}_\tau\), and metavariables \(?_{m_j}\) are solved to \(\mathit{sol}_j\). Every such substitution and the resulting core term \(\mathit{res}\) are explicitly recorded in the tree, enabling retrospective extraction of each term’s final form.

Tactic invocations produce structured proof states: when a tactic \(t\) is applied, the tree logs the transition
\[
  (N,\;t)\;\longmapsto\;(N',\;\{\mathit{goals}',\;\Delta'\},\;\mathit{res}_t),
\]
where \(N'\) is the succeeding node, \(\{\mathit{goals}',\;\Delta'\}\) the updated set of open goals and metavariable assignments, and \(\mathit{res}_t\) the resulting proof object. This makes it possible to extract, for each invocation, both the term result and the precise proof state.

Term elaboration similarly records parsing and resolution steps: an expression \(e\) is parsed to \(e_{\mathrm{raw}}\), resolved to \(e_{\mathrm{typed}}\), and finally elaborated to \(e_{\mathrm{core}}\); each intermediate form and corresponding state are stored as separate nodes, allowing fine-grained recovery of every subterm’s elaboration result and its local context.

We obtain these elaboration trees via the \texttt{leanjixia} interface, which exposes Lean’s internal elaborator. A downstream Python analysis layer then traverses the tree to extract for each node both the elaborated core term \(\mathit{res}\) and its associated proof state \((\Gamma,\Sigma,\Delta,\mathit{ctx})\). These extracted sequences of \((\mathit{expr},\mathit{res},\text{state})\) form the basis for our chain-of-states methodology, enabling systematic expansion of implicit reasoning steps into explicit intermediate states.  
For each node in the elaboration tree, we can obtain the following information: the node type, 
the variables and theorems used, and the proof state at the node. 

By analyzing this tree, we can expand common syntactic combinations. We leverage metaprogramming techniques based on Leanjixia \cite{jixia} to extract data by analyzing the elaboration trees of all Lean files in the mathematical domain. Leanjixia is an open-source static analysis tool for Lean that additionally supports dynamic Python-Lean interaction. Compared to Leandojo \cite{Yang2023Leandojo}, it offers advantages including lower memory consumption and higher parallel processing efficiency. In the proof state, we record all current goals, with each goal containing its assumptions and target. When a tactic or term is internally expanded, the unfolded context corresponds to the state before the tactic or the state containing the term. For theorem composition, we incorporate the result of the inner composition as a hypothesis in the outer proof goal, breaking complex term compositions into a sequence of simpler tactics. 

\subsection{Enhanced Chain of States Through Composite terms in Tactic and Term-Based Proof}

Using the \texttt{leanjixia} framework, we extract elaboration trees from Lean proofs to analyze and reconstruct fine-grained proof structures. Each node in the elaboration tree contains rich metadata, including the elaboration type, the corresponding character span in the source code, the in-scope variables, and the proof state (or \textit{State}) at the time of elaboration. Since tactic-based proofs in Lean are typically introduced via the keyword \texttt{by}, we collect all subtrees rooted at \texttt{by} nodes and treat them as tactic-level proof structures for downstream analysis and training.

While this strategy captures a large portion of formal proof steps, we observe that in \textit{mathlib}, many proofs involve complex composite terms—either embedded within tactics or used as standalone term-based proofs. These composite structures are often built through deeply nested applications of lemmas or functional terms. In practice, our goal is to decompose such term-based proofs into stepwise, fine-grained tactic-based representations. However, this is nontrivial due to the lack of a canonical, universally agreed granularity for tactics: determining how to break a term or compound tactic into semantically meaningful steps remains ambiguous and context-dependent.

Interestingly, the elaboration tree offers insight into the internal structure of composite terms through nodes labeled with \texttt{term.app}. During elaboration, each subterm is recursively analyzed, and its expected type is carefully recorded. For example, as shown in the first theorem in Figure~\ref{fig:lean_composite}, the proof applies the lemma \texttt{even\_x\_plus\_2} twice to reach the goal without any explicit \texttt{by} block. The inner application \texttt{even\_x\_plus\_2\ h} is elaborated with the expected type $\exists k,\ x = 2 \cdot k$, which precisely satisfies the input requirement of the outer application. Similarly, in the second theorem, the terms \texttt{sq\_nonneg\ a} and \texttt{sq\_nonneg\ b} are intermediate expressions with expected types $a^2 \ge 0$ and $b^2 \ge 0$, respectively, making implicit facts explicit.

\begin{figure}[ht]
  \centering
  \begin{tcolorbox}[
    colback=gray!10,
    colframe=black,
    boxrule=0.5pt,
    arc=1mm,
    left=2mm, right=2mm,
    top=0mm, bottom=0mm,
    width=0.9\linewidth
  ]
\begin{lstlisting}[language=Lean]
lemma even_x_plus_2 {x : Nat} 
    (h : ∃ k, x = 2 * k) : ∃ k, x + 2 = 2 * k := sorry

theorem term_based_proof
    (x : Nat) (h : ∃ k, x = 2 * k) :  ∃ k, x + 4 = 2 * k :=
      even_x_plus_2 (even_x_plus_2 h)

theorem term_in_tactic
    (a b : ℝ) : a ^ 2 + b ^ 2 ≥ 0 := by
        linarith [sq_nonneg a, sq_nonneg b]
\end{lstlisting}
  \end{tcolorbox}
  \caption{Composite Terms in Lean}
  \label{fig:lean_composite}
\end{figure}

\begin{figure}[htbp]
  \centering
  \includegraphics[width=0.6\linewidth]{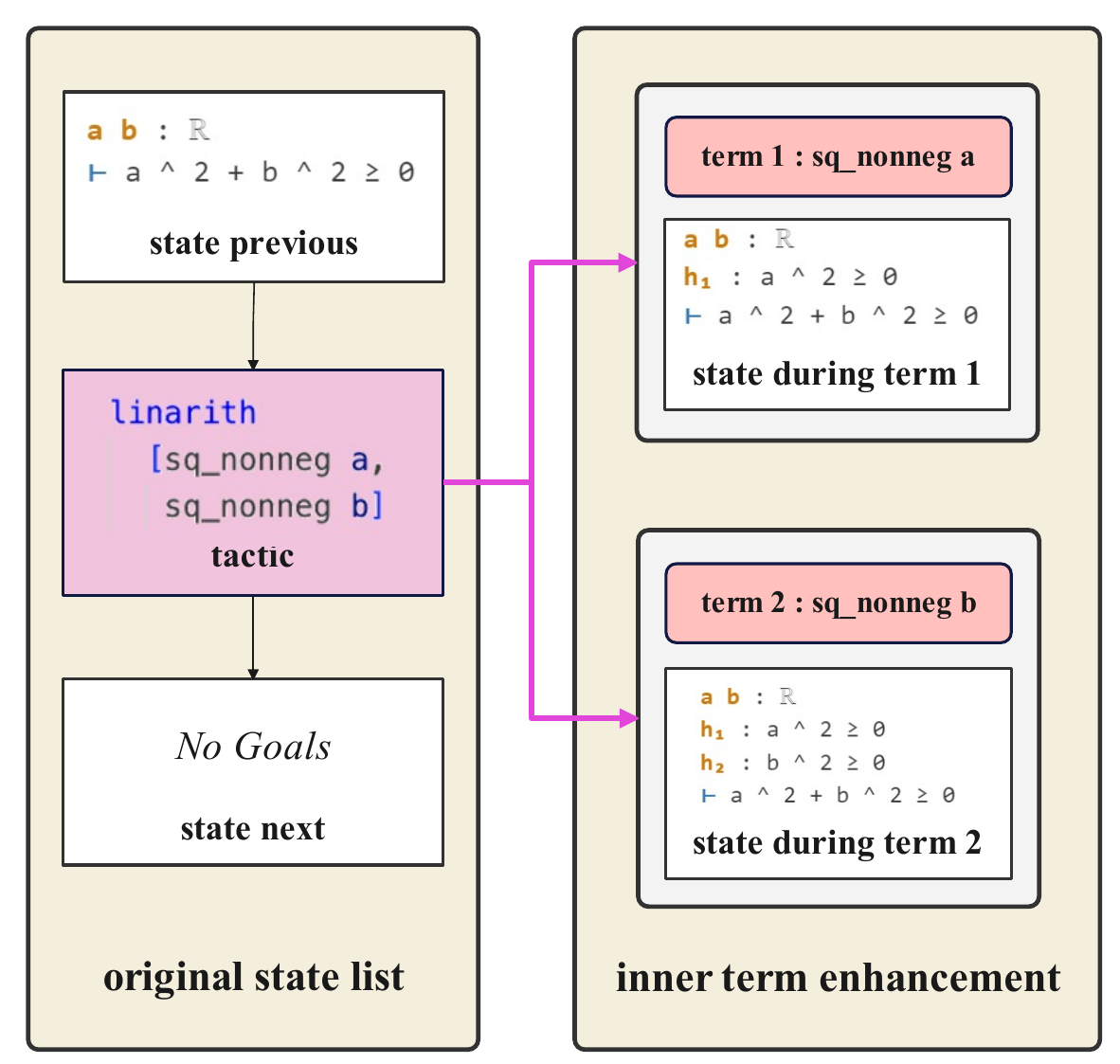}
  \caption{Enhanced CoS Through Composite Terms}
  \label{fig:cse_expr}
\end{figure}

Building on this structure, we introduce a state-enhancement mechanism for composite terms. For a composite term, we extract its elaboration state \(\mathcal{S}_p\) and traverse the elaboration tree root-to-leaf. At each internal node, we append the node’s expected type as a new hypothesis to the main goal of \(\mathcal{S}_p\). This lightweight pass inserts semantic checkpoints—intermediate facts implied by the term—and enriches the available context.
As an illustrative example, consider the second theorem in Figure~\ref{fig:lean_composite}. The original chain of states (CoS) has two nodes, $\mathcal{S}_p$ (previous) and $\mathcal{S}_n$ (next). After the enhancement, we insert two additional states, each recording the elaboration trace and the expected type of one subterm, so the chain now has four states (see Figure~\ref{fig:cse_expr}): \texttt{state previous} $\to$ \texttt{state during term 1} $\to$ \texttt{state during term 2} $\to$ \texttt{state next}.
For the first theorem, which originally uses a term-based proof without tactics, the initial CoS contains only the original state followed by a terminal \texttt{No goals}. After enhancement, we introduce an informative intermediate state with an additional hypothesis $\mathit{h}_0 : \exists\,k,\ x + 2 = 2 * k$, which mirrors the logical effect of a \texttt{have} step and clarifies intent.

This enhancement significantly alleviates the over-representation of short two-step CoS chains in the raw \textit{mathlib} dataset. Without such processing, the CoS generator model often outputs degenerate, overly terse proof trajectories that jump directly from the original state to \texttt{No goals}, failing to meaningfully decompose the problem. In contrast, enhanced chains promote richer intermediate structure and more faithful reasoning flow, leading to better learning signals and improved downstream robustness.

Furthermore, Lean proofs sometimes contain nested \texttt{by} blocks, typically within a \texttt{have} tactic. If processed naively, such nested blocks interrupt the main CoS and inject a new sub-chain into the middle of an otherwise linear proof trajectory. To handle this, we extract the inner \texttt{by} blocks as independent sub-CoS and convert their initial states into standalone formal statements. In the outer proof, each such nested proof is treated as a single tactic-level application, rather than a sequence of multiple tactic steps. This treatment preserves the coherence of the outer proof and maintains the integrity of the overall proof trajectory.

\section{Tactic Generation from Chain of States}
\label{sec :tactic generation}
\subsection{Tactic Generation from Adjacent States}

Although generating tactics between adjacent states is relatively simple for humans, it presents a novel problem paradigm for most ATP models. Conventional ATP models generate tactics from an original state, continuing until producing a “No Goals” state, timing out, or reaching maximum search depth. This global rollout often overlooks the fine structure of local transitions between adjacent proof states. For certain special cases of adjacent states, we can perform specific transformations to convert them into standard ATP problems. For example, if $\mathcal{S}*n$ differs from $\mathcal{S}*p$ only by adding a condition $h*{\text{new}}$ in the main goal, the task reduces to proving $h*{\text{new}}$ under the existing hypotheses of the main goal in $\mathcal{S}_p$. In general, however, we rely on \textbf{higher-order implications} and subtle context edits, such as quantifier instantiation, binder introduction, and hypothesis reindexing, which standard ATP objectives rarely expose.

Consider the case where both $\mathcal{S}_p$ and $\mathcal{S}_n$ contain only a single goal. Suppose the hypotheses in $\mathcal{S}_p$ and $\mathcal{S}_n$ are denoted by $h_p$ and $h_n$, and the respective goals are $\tau_p$ and $\tau_n$. Then, the problem reduces to proving the higher-order implication 
$h_p \rightarrow (h_n \rightarrow \tau_n) \rightarrow \tau_p$. Assuming its proof is given by \texttt{tactics\_new}, the adjacent states can then be transformed as follows:

\begin{tcolorbox}[
  colback=gray!10,    
  colframe=black,     
  boxrule=0.5pt,      
  arc=1mm,            
  left=4mm, right=2mm,
  top=4mm, bottom=2mm
]
\begin{lstlisting}
have t_new : (hₙ → τₙ) → τ := tactics_new
apply t_new
repeat' assumption
\end{lstlisting}
\end{tcolorbox}

However, higher-order implication transformations significantly alter state distributions, which particularly affects LLM-based automated theorem provers (ATPs). These models are typically trained on special-case subproblems and lack exposure to higher-order logic transformations, leading to suboptimal performance. Consequently, rather than relying on conventional ATP-oriented LLMs, we opt to directly provide adjacent states to general-purpose large language models. Given that most LLMs are predominantly trained on Lean3 data and exhibit limited tactic generation capability, we selected Deepseek-R1 as our model for generating tactics between adjacent states.

In addition, our custom built \textbf{common tactic prover} systematically explores standard Lean automation tactics. It offers stable, efficient performance for simple problems and remains unaffected by logical transformations. For Lean tactic generation, we first employ Leanjixia as our Python-Lean interaction framework. Within this framework, our Lean tactic prover sequentially attempts the following tactics on $\mathcal{S}_{p}$ : "nlinarith", "linarith", "aesop", "omega", "field\_simp", "ring", "simp", "norm\_num". These single-step deterministic strategies require no depth or width search. We further integrate the semi-interactive "apply?" tactic into our common tactic prover. This tactic searches through the imported theorem library, attempting "apply" tactics on the current goal and iteratively applying the same process to resulting states. After six attempts, "apply?" evaluates proof completion status, returning successful tactics or presenting intermediate states for selection. Traditional implementations either lack interactivity or misjudge proof completion in Python-Lean gyms. Through metaprogramming, we enhance Leanjixia to properly handle "apply?" operations and collect all return information, even for incomplete proofs.

\subsection{Error Tactic Regeneration and Error State Renewal}

\begin{algorithm}[H]
\label{alg: proof generation}
\caption{Formal Proof Generation via CoS}
\KwIn{Informal proof $\mathsf{inf\_proof}$, statement $\mathsf{state_0}$, CoS\_LLM, $n_{\text{CoS}}$, $n_P$}
\KwOut{CoS, formal proof}

Rewrite $\mathsf{inf\_proof}$ to Lean-friendly form via Deepseek\_R1\;
Generate $n_{\text{CoS}}$ candidates with CoS\_LLM $\rightarrow$ $\mathsf{CoS\_list}$\;
Original CoS $[\mathsf{state_0}, \texttt{proved}] \rightarrow \mathsf{CoS\_list}$\;

\ForEach{$\mathsf{CoS} \in \mathsf{CoS\_list}$}{
  \For{$i \gets 1$ \KwTo $n_P$}{
    Initialize empty $\mathsf{tactic\_list}$\;
    \ForEach{adjacent state $(s_p, s_n)$ in $\mathsf{CoS}$}{
      Try common tactic prover on $(s_p, s_n)$ $\rightarrow \mathsf{tactics}$\;
      \If{valid $\mathsf{tactics}$ leads to $s_n$ via \text{Leanjixia}}{
        $\mathsf{tactics} \rightarrow \mathsf{tactic\_list}$; \textbf{continue}\;
      }
      Generate $\mathsf{tactics}$ via DeepseekR1 on $(s_p, s_n)$ or with $\mathsf{error\_info}$ if prior fails\;
      Use Leanjixia to get $\hat{s}_n$ from $(s_p, \mathsf{tactics})$\;
      \uIf{$\hat{s}_n = s_n$}{
        $\mathsf{tactics} \rightarrow \mathsf{tactic\_list}$\;
      }\uElseIf{$s_n = \texttt{proved}$}{
        Suggest tactic via Deepseek\_R1 with special prompt and verify via Leanjixia\;
        \If{proof complete}{$\mathsf{tactics} \rightarrow \mathsf{tactic\_list}$}\Else{break}
      }\Else{
        Set $s_n \gets \texttt{proved}$; retry via Deepseek\_R1\;
        \If{proof complete}{$\mathsf{tactics} \rightarrow \mathsf{tactic\_list}$}\Else{break}
      }
    }
    \If{proof complete}{\textbf{return} $(\mathsf{CoS}, \mathsf{tactic\_list})$}
  }
}
\end{algorithm}
\label{error back}
In this section, we utilize the language model Deepseek-R1 to generate intermediate tactics given a pair of adjacent proof states. To ensure compatibility with the original ATP formulation, where the goal is to generate tactics solely from a given state $\mathcal{S}_p$, we set the subsequent state $\mathcal{S}_n$ to "no Goals". In this way, whether proving a state directly or completing proofs between adjacent states, our framework allows Deepseek-R1 to output a valid segment of tactics based on the input pair $(\mathcal{S}_p, \mathcal{S}_n)$.
To address this issue, most theorem provers offer multiple possible tactics at each state and attempt them. Invoking Deepseek-R1 to generate tactics also encounters similar problems. To handle such failures, besides retrying directly, leveraging the advantages of using a general model, we devise the \textbf{error tactic regeneration} and \textbf{error state renewal} algorithm.

\textbf{Error Tactic Regeneration}. After invoking Deepseek-R1 to generate tactics, if the current proof state $\mathcal{S}_{p}$ fails to transition successfully into the target state $\mathcal{S}_{n}$, we extract error information using Leanjixia. The errors fall into two categories: (1) syntax errors, which result from violations of Lean's syntax rules, and (2) semantic errors, where the generated state does not match the expected $\mathcal{S}_{n}$. When such errors occur, the corresponding feedback is incorporated into the prompt for Deepseek-R1, guiding it to regenerate improved tactics.

\textbf{Error State Renewal}. In testing error tactic regeneration, we find that Deepseek-R1 sometimes skips steps, i.e., the generated tactics yield a later state in the same chain. This phenomenon is common in states further down the chain or in states involving case analysis. 
We attribute this behavior primarily to the nature of Deepseek-R1’s training data, which predominantly consists of direct proof completions from $\mathcal{S}_{p}$,
and emphasizes case-analysis patterns.
Hence, if Deepseek-R1 still fails to yield $\mathcal{S}_{n}$ after error tactic regeneration, we modify $\mathcal{S}_{n}$ in the input to "No Goals", asking Deepseek-R1 to attempt proving $\mathcal{S}_{p}$ directly.

\section{Experiments}
\label{experiments}
\subsection{Experiment Settings}
\subsubsection{Chain of States Generator LLM Training Settings}

To train the CoS generator, we perform supervised fine-tuning using a distillation of the Deepseek Qwen-7B model. The training is conducted for 1 epoch on two A800 GPUs, with a learning rate set to \texttt{5e-4}. The training corpus consists of tuples of formal statements, informal proofs, detailed informal proofs, and their corresponding Chain of States representations. 

During fine-tuning, the model is guided by a task-specific system prompt designed to instruct it in generating logically coherent state sequences aligned with informal mathematical reasoning. The prompt enforces structural and semantic constraints to ensure consistency and completeness of the CoS output. The full system prompt used during training is in Appendix \ref{Prompt_COS}:

\noindent
\textbf{Post-Processing.} Since the output chain of states may include redundant or repeated states due to the model’s generation behavior, we apply a deduplication step to remove duplicated adjacent states. Additionally, we verify that the output includes a complete and valid chain—ending in a state labeled as \texttt{No Goals}—and if not, we trigger re-generation to ensure a logically complete output. This step is critical to maintain the integrity and usability of CoS sequences for downstream learning and verification tasks.

\subsubsection{Tactic Generation Settings}
\textbf{Benchmark}: Minif2f, a standard dataset for formal Lean4 proofs, consisting of two splits: test and valid. It is widely used for benchmarking mathematical reasoning and theorem-proving models. Due to version compatibility issues, we test 227 and 229 problems on the MiniF2F-test and MiniF2F-valid sets, respectively.

\textbf{Baselines}: In our experiments, we evaluated several models familiar with Lean reasoning, as well as baseline models for Lean4. The models tested are described below:

1. DeepSeek-Prover-V1.5: DeepSeek-Prover-V1.5 employs a whole-proof generation approach, trained on large-scale synthetic proof data. It combines supervised fine-tuning (SFT) and reinforcement learning (RL) to optimize proof generation and alignment with Lean's formal verification system. In our experiments, each problem was repeated 16 times to accelerate runtime.

2. InternLM2.5-StepProver: A step-by-step proving model based on InternLM2.5, designed for efficient tactic generation and proof exploration.

3. HunyuanProver: A theorem-proving model optimized for Lean4, leveraging advanced search strategies and large-scale training data.

\textbf{Tactic Budget.} For different tactic generation models, we define their generation budgets according to their underlying inference strategies. For single-pass sampling methods, we define the sample budget $K$ as the total number of proofs generated. Larger values of $K$ are factorized to facilitate comparison with tree search-based methods. For best-first search methods, following the notation introduced by Azerbayev et al.~(2024), we define $K = N \times S \times T$, where $N$ is the number of best-first-search attempts, $S$ denotes the number of tactics generated per expansion, and $T$ represents the number of expansion steps. For tree-based search methods such as RMaxTS and HTPS (Lample et al., 2022), we express $K$ as $K = N \times T$, where $N$ is the number of search restarts and $T$ is the number of tactic generation calls made during tree expansion.

In our proposed model, we empirically observe that the maximum Chain of States (CoS) length across all MiniF2F problems is 15. For each problem, we generate 20 independent CoS samples, each undergoing 10 rounds of adjacent-state tactic generation. Each round permits up to 3 additional tactic calls due to error-based regeneration and state renewal, if necessary. For the final CoS variant—where the chain is simplified by truncating to only the last state—we apply 16 tactic generation calls, with each tactic allowed up to 2 regeneration attempts in the event of errors.
\subsection{Chain of States Generation}
\label{cosgeneration}
Since the generated chain of states cannot directly produce corresponding tactics, we require an efficient method to evaluate their quality. To address this, we propose the CoS benchmark, which includes 100 problems curated from Mathlib and Lean Workbook (80 from Mathlib and 20 from Lean Workbook). To ensure the diversity of the dataset, we employ the following selection strategy. For Mathlib, we uniformly sample 80 problems across different mathematical domains, using GPT-4o to complete their informal proofs. For Lean Workbook, we randomly select 20 previously proven problems and perform the same proof completion process.

\begin{table}[ht]
\centering
\caption{Performance comparison of generating CoS across benchmark datasets.}
\begin{tabular}{lccccc}
\toprule
Method & \#Params &  Budget &Minif2f-test & Minif2f-valid & CoS benchmark \\ 
\midrule
GPT4o   & closed-source& 1& 46.31\% & 40.57\% & 29.00\% \\ 
GPT4o   & closed-source& 5& 95.90\% & 94.90\% & 40.00\% \\ 
DeepseekR1 & 600B &  1 & 16.80\% & 18.03\%& 23.00\% \\ 
DeepseekR1 & 600B &  5 & 67.62\% & 61.48\%& 44.00\% \\ 
CoSLLM & 7B &1 & 58.60\% & 55.73\% & 32.00\% \\
CoSLLM & 7B &5 & 95.49\% & 94.67\% & 53.00\% \\  
\bottomrule
\end{tabular}
\label{tab:results}
\end{table}

The evaluation of generated state chains follows a two-phase verification process. First, GPT-4o performs a semantic consistency check to verify whether the chain of states semantically aligns with the provided informal proof. Then, Leanjixia conducts a syntactic validity check to validate Lean syntax compliance. A state chain is deemed qualitatively acceptable only upon passing both verification stages. A performance comparison of the benchmark across the general-purpose model GPT-4o, Deepseek-R1, and our proposed model is shown in Table \ref{tab:results}. We find that R1's performance is significantly worse than other models because it confuses tactics with states and fails to output the correct format. On the MiniF2F benchmark, our 7B model significantly outperforms GPT-4o in single-generation (1 CoS) settings, while showing comparable performance in 5 CoS generations. Again, on the CoS benchmark, our model demonstrates superior performance over GPT-4o in both 1 CoS and 5 CoS settings, highlighting its robust semantic alignment capability across diverse mathematical domains.

\subsection{Tactic Generation}
\label{tactic_generation in experiments}
To evaluate whether translated informal proofs can correctly generate tactics, we employ a CoS generator LLM to produce multiple state chains. For each chain, we apply the tactic generation framework shown in Algorithm \ref{alg: proof generation} and conduct experiments on the Minif2f dataset.
For the data processing,  we utilize an open-source repository Minif2f-lean4 from HuggingFace containing informal proofs.
 However, we identify several issues including misaligned notation between informal and formal proofs, excessive proof skipping, and logical gaps. We manually correct these issues for a subset of the informal proofs. We next explain tactic generation protocol.  
For the first 19 state chains (containing intermediate states), we perform 10 tactic generation attempts per adjacent state pair. Each failed generation triggers one round of error tactic regeneration and error state renewal. For the 20th chain (without intermediate states), we conduct 16 initial tactic generations followed by error tactic regeneration if needed, maintaining comparable search effort. This protocol yields a maximum of $20 \times \text{len(CoS)} \times 32$ Deepseek-R1 calls per problem. After deduplication, the maximum chain length (len(CoS)) in Minif2f is 15, with average length around 4.

\begin{table}[h]
\centering
\caption{Comparison of Proof Generation and Tree Search Methods}
\begin{tabular}{l l l l}
\toprule
\textbf{Method} & \textbf{Budget} & \textbf{Minif2f-Valid} & \textbf{Minif2f-Test} \\
\midrule
TheoremLlama \citep{wang2024theoremllama} & cumulative & 36.5\% & 33.6\% \\
Deepseek-Prover \citep{xin2024deepseekv1} & 16$\times$4096 & - & 50.0\% \\
ReProver \cite{Yang2023Leandojo} & - & - & 26.5\% \\
Llemma-34B \citep{azerbayev2023llemma} & 1$\times$32$\times$100 & 27.9\% & 25.8\% \\
Curriculum Learning \citep{polu2022formal} & 64$\times$8$\times$512 & 47.3\% & 36.6\% \\
Lean-STaR \citep{lin2024lean} & 64$\times$1$\times$50 & - & 46.3\% \\
Deepseek-Prover-V1.5-RL \citep{xin2024deepseek} & 32$\times$6400 & - & 63.5\% \\
Goedel-Prover \citep{goedelprover} & 4$\times$6400 & - & 64.7\% \\
InternLM2-StepProver \citep{wu2024lean} & 64$\times$32$\times$100 & 63.9\% & 54.5\% \\
InternLM2.5-StepProver-BF+CG \citep{wu2024internlm2} & 256$\times$32$\times$600 & 69.6\% & 65.9\% \\
HunyuanProver v16+BFS+DC \citep{li2024hunyuanprover} & 600$\times$8$\times$400 & - & 68.4\% \\
BFS-Prover \citep{xin2025bfs} & 512$\times$2$\times$600 & - & 68.7\% \\
CoSProver& \textbf{20$\times$15$\times$32} & \textbf{70.3\%} & \textbf{69.2\%} \\
\bottomrule
\end{tabular}
\label{table:proof_generation_comparison}
\end{table}

We present separate results in Tables \ref{table:proof_generation_comparison} for Minif2f-test and Minif2f-valid. Our method achieves state-of-the-art performance on both datasets while operating in a significantly reduced search space, demonstrating the superiority of our approach in terms of efficiency and effectiveness. In particular, while prior methods often rely on thousands to tens of thousands of tactic generation calls per problem, our approach achieves higher proof success rates with only a few hundred calls—representing a 2–3 orders of magnitude reduction in computation. This highlights not only the accuracy but also the practicality of our framework, making it suitable for real-world settings where computational resources are limited.

\subsection{Ablation}
\label{ablation}

We isolate the contribution of each component in \textsc{CoSProver} without additional subpoints. The R1-only baseline (32 invocations) is modest. Adding error tactic regeneration (ETR) yields a sizeable jump, confirming the efficacy of targeted local repair. Enabling Chain-of-State (CoS) exploration together with error state renewal (ESR) further improves correctness via breadth-first state coverage and structured retries. The full system under the $20\times15\times32$ budget reaches $69.16\%$ on MiniF2F-Test. In practice, Lean-friendly rewriting of informal proofs (Sec.~\ref{leanfriend}) synergizes with these modules by reducing brittle edges in tactic synthesis and stabilizing downstream search.

\begin{table}[h]
  \centering
  \caption{Ablation on \textsc{CoSProver} modules. ``ETR'': error tactic regeneration; ``ESR'': error state renewal. The $a\times b\times c$ budget denotes \#CoS candidates $\times$ max adjacent state pairs per CoS $\times$ max tries per local tactic.}
  \label{tab:ablation}
  \begin{tabular}{lcc}
    \toprule
    Prover System & Tactic Budget & MiniF2F-Test \\
    \midrule
    CommonTacticProver & -- & 29.52\% \\
    DeepSeek\textendash R1 & $32$ & 37.00\% \\
    DeepSeek\textendash R1 + ETR & $32$ & 49.78\% \\
    DeepSeek\textendash R1 + 10 CoS + ETR + ESR & $10\times15\times30$ & 58.59\% \\
    \textbf{CoSProver (full)} & $20\times15\times32$ & \textbf{69.16\%} \\
    \bottomrule
  \end{tabular}
\end{table}

\begin{figure}[h]
  \centering
  \includegraphics[width=0.55\textwidth]{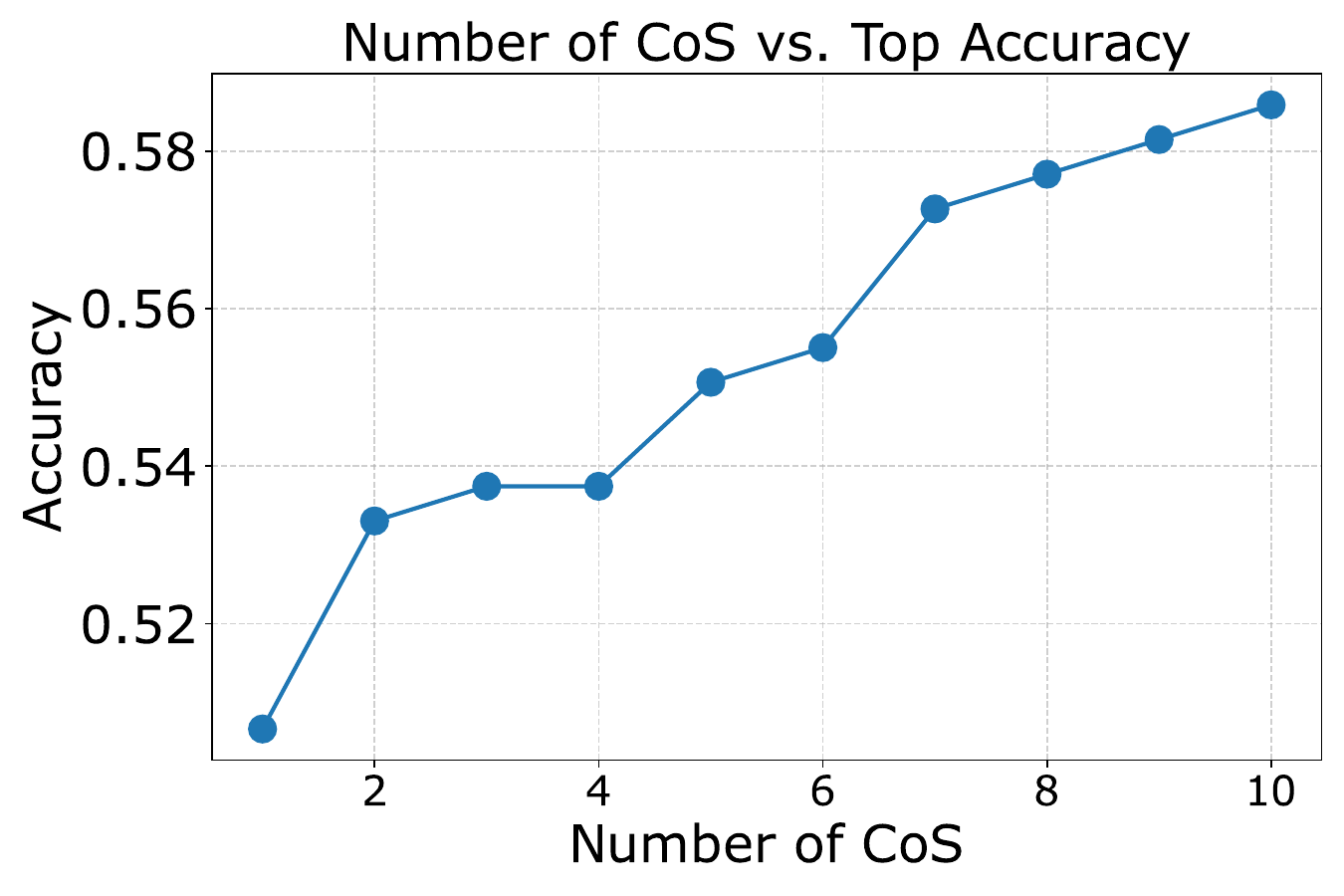}
  \caption{Accuracy vs.\ CoS iteration} 
  \label{fig:cos_depth}
\end{figure}

\textit{Depth study on CoS iterations.}
Holding the execution counts of direct proving, ETR, and ESR fixed at $10$, we vary the number of CoS iterations using sequences extracted \emph{directly} from unmodified informal proofs to avoid confounds, with all other settings (model, dataset, and tactic budgets) kept constant. Accuracy on MiniF2F-Test increases steadily with depth (Figure~\ref{fig:cos_depth}): $5$ iterations already outperform the baseline, and $10$ iterations reach $58.59\%$ in this controlled setup. Beyond $10$, we observe diminishing yet consistent gains. These improvements arise from broader state-space coverage and more opportunities to repair local failures, rather than from ancillary heuristics.

\textit{Comparison with direct generation and translation.}
To provide a clear reference point, we compare two non-agentic baselines under matched call budgets: end-to-end direct proof generation with DeepSeek--R1 and end-to-end NL$\to$Lean translation, including a variant that applies two ETR passes after a single translation attempt. These comparisons make explicit that, relative to using a strong model purely in generation/translation modes, our structure-aware \textsc{CoS} pipeline attains higher accuracy via state decomposition, ETR, and ESR---rather than by increasing the number of calls. Note that our main experiments use a $20\times15\times32$ tactic budget (20 candidate \textsc{CoS} paths $\times$ up to 15 adjacent-state pairs/subproblems per path $\times$ up to 32 tries per local tactic). Here we report 32--192 calls to align with the per-subproblem attempt scale: \textsc{CoS} decomposes a theorem into up to 15 subproblems and explores up to 20 proof paths, but each subproblem receives at most 32 local attempts; we therefore sweep $\{32, 64, 128, 192\}$ as multiplicative single-subproblem budgets for a fair and interpretable comparison.

\begin{table}[h]
  \centering
  \caption{Direct generation vs.\ NL$\rightarrow$Lean translation under matched call budgets on MiniF2F-Test (\%).
  ``ETR'': two extra ETR passes executed after one translation pass.}
  \label{tab:comp_gen_trans_row}
  \begin{tabular}{lcc}
    \toprule
    Prover System & Tactic Budget & MiniF2F-Test \\
    \midrule
    Direct generation & $32$ & 35.2\% \\
    Direct generation & $64$ & 40.5\% \\
    Direct generation & $128$ & 43.2\% \\
    Direct generation & $192$ & 45.8\% \\
    Translation (no ETR) & $32$ & 34.8\% \\
    Translation (no ETR) & $64$ & 38.7\% \\
    Translation (+ ETR; 2 passes) & $32$ & 37.4\% \\
    Translation (+ ETR; 2 passes) & $64$ & 41.0\% \\
    \bottomrule
  \end{tabular}
\end{table}

\textit{Lean-friendly rewriting module.}
To assess alignment effects independently of agent structure, we conduct a controlled study on MiniF2F-Test, focusing on a Lean-friendly rewriting component that converts mathematically valid but structurally loose proofs into Lean-aligned text using type-specific few-shot prompts (see Section~\ref{leanfriend} and the recurring differences summarized in Table~\ref{tab:differences}). We run controls with the \emph{original} (unprocessed) proofs to isolate the effect of rewriting while keeping call/tactic budgets matched. As summarized in Table~\ref{tab:rewrite_control_inline}, \textsc{CoSProver} remains effective without rewriting; however, the rewriting module yields substantial gains at larger budgets and task scale. Mechanistically, rewriting strengthens semantic and structural compatibility between informal reasoning and formal tactics by making case splits, rewrites, calculational hints, and type/domain annotations explicit, thereby reducing tactic brittleness and stabilizing downstream search. These improvements complement the benefits of CoS, ETR, and ESR rather than relying on more calls (cf.\ Section~\ref{ablation}).

\begin{table}[h]
  \centering
  \caption{Control using \emph{original} (unprocessed) proofs at two tactic budgets}
  \label{tab:rewrite_control_inline}
  \begin{tabular}{lcc}
    \toprule
    Prover System & Tactic Budget & MiniF2F-Test \\
    \midrule
    Original proofs (no rewriting) & $1\times15\times30$  & 50.7\% \\
    Original proofs (no rewriting) & $20\times15\times30$ & 62.1\% \\
    \bottomrule
  \end{tabular}
\end{table}

\subsection{Case Study}
\subsubsection{Example of CoSProver with no Error Back}
\begin{tcolorbox}[
  colback=gray!10,    
  colframe=black,     
  boxrule=0.5pt,      
  arc=1mm,            
  left=2mm, right=2mm,
  top=0.2mm, bottom=0.2mm
]\begin{lstlisting}[language=Lean]
theorem numbertheory_x5neqy2p4 (x y : ℤ) : x^5 ≠ y^2 + 4 := by
-- x y : ℤ
-- ⊢ x ^ 5 ≠ y ^ 2 + 4
  have h_0 : x ^ 5 % 11 = 0 ∨ x ^ 5 % 11 = 1 ∨ x ^ 5 % 11 = 10 := by
    have : x % 11 = 0 ∨ x % 11 = 1 ∨ x % 11 = 2 ∨ x % 11 = 3 ∨ x % 11 = 4 ∨ x % 11 = 5 ∨ x % 11 = 6 ∨ x % 11 = 7 ∨ x % 11 = 8 ∨ x % 11 = 9 ∨ x % 11 = 10 := by omega
    rcases this with (h₃ | h₃ | h₃ | h₃ | h₃ | h₃ | h₃ | h₃ | h₃ | h₃ | h₃) <;> simp [h₃, pow_succ, Int.mul_emod, Int.add_emod]
-- x y : ℤ
-- h_0 : x ^ 5 % 11 = 0 ∨ x ^ 5 % 11 = 1 ∨ x ^ 5 % 11 = 10
-- ⊢ x ^ 5 ≠ y ^ 2 + 4
  have h_1 : y ^ 2 % 11 = 0 ∨ y ^ 2 % 11 = 1 ∨ y ^ 2 % 11 = 3 ∨ y ^ 2 % 11 = 4 ∨ y ^ 2 % 11 = 5 ∨ y ^ 2 % 11 = 9 := by
    have : y % 11 = 0 ∨ y % 11 = 1 ∨ y % 11 = 2 ∨ y % 11 = 3 ∨ y % 11 = 4 ∨ y % 11 = 5 ∨ y % 11 = 6 ∨ y % 11 = 7 ∨ y % 11 = 8 ∨ y % 11 = 9 ∨ y % 11 = 10 := by omega
    rcases this with (h₄ | h₄ | h₄ | h₄ | h₄ | h₄ | h₄ | h₄ | h₄ | h₄ | h₄) <;> simp [h₄, pow_succ, Int.mul_emod, Int.add_emod]
-- x y : ℤ
-- h_0 : x ^ 5 % 11 = 0 ∨ x ^ 5 % 11 = 1 ∨ x ^ 5 % 11 = 10
-- h_1 : y ^ 2 % 11 = 0 ∨ y ^ 2 % 11 = 1 ∨ y ^ 2 % 11 = 3 ∨ y ^ 2 % 11 = 4 ∨ y ^ 2 % 11 = 5 ∨ y ^ 2 % 11 = 9
-- ⊢ x ^ 5 ≠ y ^ 2 + 4
  omega -- No Goals
\end{lstlisting}
\end{tcolorbox}

\begin{tcolorbox}[
  colback=gray!10,    
  colframe=black,     
  boxrule=0.5pt,      
  arc=1mm,            
  left=2mm, right=2mm,
  top=2mm, bottom=2mm
]
informal proof: \\
We have that the remainder of the division of $x^5$ by $11$ is either $0$, $1$, or $10$: 
$x^5 \equiv k \mod 11$, with $k \in \{0, 1, -1\}$.\\
Similarly, we observe that $y^2 \equiv k \mod 11$, with $k \in \{0, 1, 3, 4, 5, -2\}$.\\
So $y^2 + 4 \equiv k \mod 11$, with $k \in \{4, 5, 7, 8, 9, 2\}$.\\
As a result, $x^5$ and $y^2+4$ have different remainders when divided by $11$, and cannot be equal.
\end{tcolorbox}
In this example, our model generate the CoS of the difficult number theory problem. Then we call Deepseek-R1 to complete the proof between adjacent states. R1 successfully complete the transformation and finish the proof.

\begin{tcolorbox}[
  breakable,
  colback=gray!10,    
  colframe=black,     
  boxrule=0.5pt,      
  arc=1mm,            
  left=2mm, right=2mm,
  top=0.2mm, bottom=0.2mm
]\begin{lstlisting}[language=Lean]
theorem mathd_numbertheory_277
  (m n : ℕ)
  (h₀ : Nat.gcd m n = 6)
  (h₁ : Nat.lcm m n = 126) :
  60 ≤ m + n := by
  by_contra h
-- m n : ℕ
-- h₀ : m.gcd n = 6
-- h₁ : m.lcm n = 126
-- h : ¬60 ≤ m + n
-- ⊢ False
  have h₂ : m < 60 := by linarith
  have h₃ : n < 60 := by linarith
-- m n : ℕ
-- h₀ : m.gcd n = 6
-- h₁ : m.lcm n = 126
-- h : ¬60 ≤ m + n
-- h₂ : m < 60
-- h₃ : n < 60
-- ⊢ False
  interval_cases m <;> interval_cases n <;>
  norm_num [Nat.gcd_eq_right, Nat.gcd_eq_left] at h₁ h₀ ⊢ <;>
  linarith -- No Goals
\end{lstlisting}
\end{tcolorbox}

This example's original natural language proof focused on analyzing the number-theoretic properties of \( m \) and \( n \), but suffered from serious step-skipping issues. We restructured the proof using the method described in \ref{exhaustion}. The revised informal proof is as follows. Based on this new informal proof, we successfully generated the CoS and the corresponding formal proof.

\begin{tcolorbox}[
  colback=gray!10,    
  colframe=black,     
  boxrule=0.5pt,      
  arc=1mm,            
  left=2mm, right=2mm,
  top=2mm, bottom=2mm
]
Lean-friendly informal proof: \\
By contradiction. \\
Suppose \( m + n < 60 \). \\
We now determine a finite and concrete range for the values of \( m \) and \( n \). Since \( m + n < 60 \), both \( m \) and \( n \) are less than 60, i.e., \( m < 60 \) and \( n < 60 \).

Furthermore, the assumptions \( \mathrm{Nat.gcd}(m, n) = 6 \) and \( \mathrm{Nat.lcm}(m, n) = 126 \) impose strict arithmetic conditions. However, instead of trying to derive properties using number-theoretic formulas, we let Lean enumerate all pairs \( (m, n) \) such that:
\[
m < 60,\quad n < 60
\]
\[
\mathrm{Nat.gcd}(m, n) = 6,\quad \mathrm{Nat.lcm}(m, n) = 126
\]

If any such pair exists with \( m + n < 60 \), it would serve as a counterexample. But enumerating all such pairs reveals that no such pair satisfies all the conditions with \( m + n < 60 \). Therefore, a contradiction arises. Hence, we conclude \( m + n \ge 60 \).
\end{tcolorbox}

\subsubsection{Example of Error State Renewal}

\begin{tcolorbox}[
  colback=gray!10,    
  colframe=black,     
  boxrule=0.5pt,      
  arc=1mm,            
  left=2mm, right=2mm,
  top=0.2mm, bottom=0.2mm
]\begin{lstlisting}[language=Lean]
theorem induction_1pxpownlt1pnx (x : ℝ) (n : ℕ) (h₀ : -1 < x) (h₁ : 0 < n) : (1 + ↑n*x) ≤ (1 + x)^(n:ℕ) := by sorry
\end{lstlisting}
\end{tcolorbox}

\begin{tcolorbox}[
  colback=gray!10,    
  colframe=black,     
  boxrule=0.5pt,      
  arc=1mm,            
  left=2mm, right=2mm,
  top=2mm, bottom=2mm
]
informal proof:\\
We show the result by induction on $n$. The result is trivial for $n=0$ or $n=1$. Let us assume the property is true in $n$.\\
By the induction hypothesis we know that $(1+nx)\leq (1+x)^n$.\\
Moreover, as $x > -1$, we have that $x \leq x (1 + x)^n$. The inequality is trivial if $x \geq 0$, and is also true if $x < 0$ as $-1 < x < 0 \implies 0 < (1 + x)^n < 1$.\\
So, $(1+nx) + x \leq (1+x)^n + x (1+x)^n$ and we have that $(1+(n+1)x) \leq (1+x)^(n+1)$, so the property is true in $n+1$.\\
By induction, we have that the result is true for any natural number $n$.
\end{tcolorbox}
At first, our model generates the following wrong intermediate state. Our model misunderstands the meaning of induction and generates the CoS only taking the base case into account. Therefore, we use the state renewal module to fix the issue. Finally, we provide the following correct proof.

\begin{tcolorbox}[
  colback=gray!10,    
  colframe=black,     
  boxrule=0.5pt,      
  arc=1mm,            
  left=2mm, right=2mm,
  top=0.2mm, bottom=0.2mm
]\begin{lstlisting}[language=Lean]
State 1:
x : ℝ
n : ℕ
h₀ : -1 < x
h₁ : 0 < n
h_0 : 1 + 0 * x ≤ (1 + x) ^ 0
⊢ 1 + ↑n * x ≤ (1 + x) ^ n
\end{lstlisting}
\end{tcolorbox}

\begin{tcolorbox}[
  breakable,
  colback=gray!10,    
  colframe=black,     
  boxrule=0.5pt,      
  arc=1mm,            
  left=2mm, right=2mm,
  top=0.2mm, bottom=0.2mm
]\begin{lstlisting}[language=Lean]
theorem induction_1pxpownlt1pnx (x : ℝ) (n : ℕ) (h₀ : -1 < x) (h₁ : 0 < n) : (1 + ↑n*x) ≤ (1 + x)^(n:ℕ) := by
  induction' h₁ with n hn
  simp_all [Nat.cast_zero, zero_mul, add_zero, pow_zero]
  simp_all [Nat.cast_succ, add_mul, mul_add, pow_succ]
  nlinarith [sq_nonneg (x - 1), sq_nonneg (x + 1), hn, add_nonneg (sq_nonneg x) (sq_nonneg 1)]

\end{lstlisting}
\end{tcolorbox}
\subsubsection{Example of Error Tactic Regeneration}
\begin{tcolorbox}[
  colback=gray!10,    
  colframe=black,     
  boxrule=0.5pt,      
  arc=1mm,            
  left=2mm, right=2mm,
  top=0.2mm, bottom=0.2mm
]\begin{lstlisting}[language=Lean,
  mathescape=true,
  columns=fullflexible, keepspaces=true, showstringspaces=false,
  literate={∑}{{$\sum$}}1 {Σ}{{$\sum$}}1 {ℕ}{{$\mathbb{N}$}}1
           {₀}{{\(_{0}\)}}1 {₁}{{\(_{1}\)}}1 {₂}{{\(_{2}\)}}1 {₃}{{\(_{3}\)}}1 {₄}{{\(_{4}\)}}1
           {₅}{{\(_{5}\)}}1 {₆}{{\(_{6}\)}}1 {₇}{{\(_{7}\)}}1 {₈}{{\(_{8}\)}}1 {₉}{{\(_{9}\)}}1
           {∀}{{$\forall$}}1 {∈}{{$\in$}}1 {↔}{{$\leftrightarrow$}}1
           {·}{{$\cdot$}}1]

theorem mathd_algebra_196 (S : Finset ℝ) (h₀ : ∀ (x : ℝ), x ∈ S ↔ abs (2 - x) = 3) : ∑ k in S, k = 4 := by
  have hS : S = { -1, 5 } := by
    ext x  
    rw [mem_insert, mem_singleton]!\textcolor{red}{----unknown identifier 'mem\_insert'}!
    apply h₀  
    rw [abs_eq_iff]  
    norm_num  
    constructor
    · intro h
      cases h with
      | inl h1 => left; linarith
      | inr h2 => right; linarith
    · intro h
      cases h with
      | inl h1 => left; rw [h1]; norm_num
      | inr h2 => right; rw [h2]; norm_num
  rw [hS]  
  simp
  norm_num  
\end{lstlisting}
\end{tcolorbox}

\begin{tcolorbox}[
  colback=gray!10,    
  colframe=black,     
  boxrule=0.5pt,      
  arc=1mm,            
  left=2mm, right=2mm,
  top=2mm, bottom=2mm
]
informal proof: \\
In order to have $|2-x| = 3$, we must have $2-x = 3$ or $2-x = -3$.  If $2-x = 3$, then $x=-1$, and if $2-x = -3$, then $x = 5$.  The sum of these solutions is $(-1) + 5 = 4$.
\end{tcolorbox}

In this example, R1 generates an incorrect proof due to an importing error. We input the error message into the ETR module and obtained a correct proof as follows:
\begin{tcolorbox}[
  breakable,
  colback=gray!10,    
  colframe=black,     
  boxrule=0.5pt,      
  arc=1mm,            
  left=2mm, right=2mm,
  top=0.2mm, bottom=0.2mm
]\begin{lstlisting}[language=Lean,
  mathescape=true,
  columns=fullflexible, keepspaces=true, showstringspaces=false,
  literate={∑}{{$\sum$}}1 {Σ}{{$\sum$}}1 {ℕ}{{$\mathbb{N}$}}1
           {₀}{{\(_{0}\)}}1 {₁}{{\(_{1}\)}}1 {₂}{{\(_{2}\)}}1 {₃}{{\(_{3}\)}}1 {₄}{{\(_{4}\)}}1
           {₅}{{\(_{5}\)}}1 {₆}{{\(_{6}\)}}1 {₇}{{\(_{7}\)}}1 {₈}{{\(_{8}\)}}1 {₉}{{\(_{9}\)}}1
           {∀}{{$\forall$}}1 {∈}{{$\in$}}1 {↔}{{$\leftrightarrow$}}1]
theorem mathd_algebra_196 (S : Finset ℝ) (h₀ : ∀ (x : ℝ), x ∈ S ↔ abs (2 - x) = 3) : ∑ k in S, k = 4 := by
  have h₁ : S = {5, -1} := by
    ext x
    simp only [Finset.mem_insert, Finset.mem_singleton, h₀, abs_eq_abs, sub_eq_iff_eq_add]
    constructor
    · intro h
      rcases eq_or_eq_neg_of_abs_eq h with (h | h)
      · right
        linarith
      · left
        linarith
    · intro h
      rcases h with (rfl | rfl)
      · norm_num
      · norm_num
  rw [h₁]
  norm_num

\end{lstlisting}
\end{tcolorbox}

\section{Conclusion}
\label{conclusion}
In this paper, we present CoSProver, a novel framework for translating informal mathematical proofs into formal Lean proofs through chain-of-states generation. Our approach begins by training a meta-learning based LLM to decompose term-based proofs into semantically meaningful intermediate states that align with informal reasoning, effectively bridging the gap between informal arguments and formal representations. To handle proof gaps between adjacent states, we develop an error recovery mechanism using Deepseek-R1 that combines both tactic regeneration and state renewal capabilities. Furthermore, we enhance the Minif2f dataset by carefully rewriting informal proofs to eliminate logical gaps while preserving their mathematical essence, and strategically incorporating tactical hints for problems that admit more Lean-friendly solutions. Experimental results demonstrate that this chain-of-states methodology substantially improves formal proof generation efficiency while maintaining a compact search space, offering a practical solution to the informal-to-formal proof translation challenge.

\begin{appendices}

\section{Prompts and Datasets in the Algorithm}

\subsection{Alignment of Informal-Formal Chain of States Datasets}

To support SFT of informal-to-formal reasoning alignment, we construct a paired dataset of informal CoS aligned with formal proof traces. Starting from the formal corpora of \texttt{mathlib}, \texttt{Leanworkbook}, and \texttt{Leanworkbook-plus}, we extract formal CoS using \texttt{leanjixia}, a metaprogramming tool that parses elaboration trees in Lean to recover detailed state transitions and applied tactics. Each CoS reflects a linearized proof trajectory consisting of a sequence of formal proof states and the corresponding tactic steps, faithfully mirroring the interactive theorem proving process.

To construct the informal counterpart of these chains, we enhance each formal CoS with three levels of natural language information: (1) informal translations of each formal proof state, (2) informal explanations of the logical transformations between adjacent states, and (3) line-by-line and full-sentence informal proofs that align with the structure of the original tactic-based proof. The informal data is generated automatically using GPT-4o with a set of structured prompts designed to preserve mathematical rigor, readability, and semantic accuracy. To avoid trivial patterns that may mislead the language model, we additionally filter out all CoS with only two states—typically consisting of an initial goal and a final \texttt{No Goals} state—since such short chains often lack sufficient structure for meaningful decomposition and lead to degenerate completions by the model.

The following prompt is used to generate the informal components for each theorem and its Lean proof script. It instructs the model to translate each state and tactic step into fluent mathematical language and reconstruct the entire informal proof in a natural and human-readable form, suitable for pedagogical and research-oriented use.

\vspace{2mm}
\noindent
\textbf{Prompt for Informal Proof, Informal State and Informal Explanation of State Change}
\begin{tcolorbox}[
  breakable,
  colback=gray!5,
  colframe=black!50,
  boxrule=0.4pt,
  arc=1mm,
  left=2mm, right=2mm, top=2mm, bottom=2mm
]
\small
Suppose you are an expert mathematician and an expert in Lean and Mathlib. First, generate the corresponding informal state based on each step of the formal state. The informal state must include all the mathematical proof information from the formal state and be narrated in natural language with a mathematician's tone. Then, your task is to translate line by line each line of the formal proof in Lean provided below into a step of informal proof. The informal proof must express the precise logic of the formal proof, written in the language of mathematicians completely, and use LaTeX. You will be provided with auxiliary information to improve the translation. Finally, please generate the whole informal proof of the theorem. Utilize the lines of informal proof written in the first task and reorganize the structure of the proof without damaging the logic. Make sure you follow the principles of whole informal proof when you generate the whole informal proof of the theorem.\\

Auxiliary Helpful Information: There are six parts of information to improve the translation. Utilize these information when generating both of the informal proof step of one line and the whole informal proof.\\

Information of the whole proof: The formal theorem, which is the goal of the whole proof. It is written in Lean. Similar translation exemplary proofs. They are written by human and attached information of the exemplary proofs and of high quality. You should learn their way of using the information and follow their style of informalization. Docstrings of the formal theorem. In most cases, the docstrings contains the informal explanation of the formal theorem, written by human. Sometimes the docstring also contains implementation notes.\\

Information of each formal proof line: When the formal proof line uses exactly one tactic, you will be provided with: Dependent definitions or statements used in the proof step. Tactic used in the formal proof line, and its explanation. Tactics are used to modify the goal or the proof context. This information is here to help you understand the logic behind each proof step. The state prior and after each line of the proof. The difference between state prior and after is exactly what the formal proof line do. When the formal proof line is a combination of more than one tactic, you will be provided with all of the above three part of information for each tactic used in the proof if necessary. Keep in mind that you should translate the entire line of formal proof to one informal proof step that contains intermediate information. DO NOT translate each tactic in one line separately into multiple steps. Utilize these information to better understand the formal proof. Although many formal details are provided, DO NOT explain any of the formal details of a proof step, or write phrases or formulas in formal style during your translation. Your task is to translate the proof steps into pure natural language without any formal influences. Make sure you follow the principles of informal proof step when you translate the formal proof into informal proof.\\

Principles of informal proof step: Express precisely and explicitly the logic structure used in each line of informal proof. If the proof step modifies or splits the goal, the logic is that "To prove some proposition (or construct something), after doing some proof steps as indicated in the tactic, it suffices to prove one or more new propositions (or construct something)." If the proof step modifies an assumption in the proof context, the logic is that "The one or more new propositions (or constructions) in the assumption is a logical consequence of the proof step (or is constructed as in the proof step)." The informal proof should be written in human-used mathematical notations and formulas in LaTeX as much as possible, and explain more detailed mathematical setup only if the definition appeared in the statement is not commonly accepted. DO NOT use to quote anything in the natural language translation, use LaTeX style and to quote. DO NOT use any of formal theorem name in the translation. Always explain the meaning of a theorem. DO NOT use h or any other formal variables to name a hypothesis. Use natural way to mention a theorem or a known hypothesis, write their meaning out explicitly. When a formal definition takes data as well as a proposition as input, do not translate the proposition in the notation in natural language. DO NOT use the word "tactic" in your translation. DO NOT use the tactics names "rewrite", "simp", ... either. Translate them into words that one use in the mathematical literature. Ensure that the translated proof maintains mathematical rigor. The informal proof should be as concise and easy to understand as possible while maintaining rigor. Keep every logical relation translated precisely inside a formal proof step. Avoid using "and" when there is some further logical relations. Rephrase the words in the proof in the customary and natural way of mathematicians. If a concept appearing in the formal proof is not in accordance with the customary usage of mathematicians, and it is explained in the auxiliary information, use this explanation to replace the concept in the informal proof. Words that are not in natural language such as Set.OrdConnected should be explained with the help of naming conventions below and the dependencies, instead of directly used. DO NOT miss any important assumptions when translation a usage of a theorem in the formal proof line. If you are not sure about the importance, write out all of them when you create the informal proof. DO NOT capital the initials of mathematical terms inside a sentence unless it is a person's name. DO NOT use the word "coe" or "coercion" or "↑". Using the help of dependency and docstring, write out the concrete coercion function explicitly. DO NOT mention anything about universe or Sort in the translation text or formulas unless this is a proof line ad hoc for universe lifting and is not a mathematical proof. DO NOT use universe-polymorphic notation .{u1,u2} in the notation. When a proof step is dealing with type coercion or anything problem that only appears in formalization, but not in mathematical proofs, DO NOT translate its formal behaviour directly. Translate it into natural mathematical language. DO NOT simply stack the states and the tactic together, show the relationship between the states and the proof in the informal translation. Explain tactic's function in natural mathematical language. DO NOT directly use their name in the translation. DO NOT explain formalization details. Provide a faithful translation of the formal proof. DO NOT add or distort information.\\

Principles of whole informal proof: It is better to restate proof in the natural order of human language, rather than the order in the formal proof. Do not damage the logic of the proof. Convert the mathematical symbols and other expressions used in the formal proof into symbols and expressions that conform to human usage. If there are formulas written in formal style, convert them to natural LaTeX style or natural mathematical language. Be as concise as possible on this basis. Convert plain text to mathematical formulas as much as possible. Especially for equalities or inequalities. Whenever possible, use chain equalities or chain inequalities. DO NOT use any formal theorem name. DO NOT use the word "tactic" or any tactic name that is not commonly used in natural language. Rephrase any word that still has some formal style left in it into completely natural mathematical language. DO NOT use the word "coe" or "coercion" or "↑". Rephrase them to their concrete meaning.
\end{tcolorbox}

\subsection{Prompt for Semantic-Syntactic Validation}

To compare the quality of CoS across different models, we apply a consistent generation prompt to general-purpose language models such as GPT-4o and Deepseek-R1. These models are given the informal proof and asked to produce a sequence of intermediate formal states representing the underlying reasoning. This setup enables a model-agnostic evaluation framework. The prompt used for this purpose is shown below.

\vspace{2mm}
\noindent
\textbf{Prompt for Generating Chain of States (GPT-4o / Deepseek-R1)}
\begin{tcolorbox}[
  breakable,
  colback=gray!5,
  colframe=black!50,
  boxrule=0.4pt,
  arc=1mm,
  left=2mm, right=2mm, top=2mm, bottom=2mm
]
\small
Assume you are an expert in mathematics and Lean. Your task is to read the following formal statement and informal proof, and generate a formal chain of states for Lean.

When generating a formal chain of states, you should adhere to the following principles:
Principles of chain of states:
1. The chain of states is a list of formal states, each state being a formal inference step in Lean. "Different states are separated by two newline characters. An example of a state is:
\begin{align*}
\text{State 0:} & \quad a : \mathbb{R},\ b : \mathbb{R},\ c : \mathbb{R},\ d : \mathbb{R},\ e : \mathbb{R},\ h : a \leq b \ \vdash\ \operatorname{rexp}(a) \leq \operatorname{rexp}(b)
\end{align*}
2. Please refer to the informal proof to generate the chain of states. Each state in your chain does not necessarily need to correspond one-to-one with each step in the informal proof. If necessary, appropriate steps required for the formal proof should be added in the output chain of states.

\end{tcolorbox}

To assess whether a generated CoS is logically consistent with the informal proof, we introduce a semantic validation stage. Given a candidate CoS, GPT-4o evaluates whether each transition step preserves the semantic content of the original reasoning and ensures alignment with the stated goal. The prompt instructs the model to perform this verification as a judgment task over the entire state sequence and informal input. The semantic validation prompt is presented below.

\textbf{Prompt for Semantic Validation (GPT-4o)}
\begin{tcolorbox}[
  breakable,
  colback=gray!5,
  colframe=black!50,
  boxrule=0.4pt,
  arc=1mm,
  left=2mm, right=2mm, top=2mm, bottom=2mm
]
\small
You are an expert in mathematics and Lean. First, check if the formal statement and State0 match exactly in meaning and variables.

If there are any new variables or unmatched letters in State0 compared to formal statement, or if they do not match in meaning, then output @@@RESULT@@@False@@@ and stop. 

Otherwise, for each formal state in the following Chain of State, first translate it into a concise informal (natural language) state. 

Then, check if these informal states, in order, are all present and described in the informal proof below, and that every reasoning step in the informal proof has its state described. If not, then output @@@RESULT@@@False@@@ and stop.

You must also check that every intermediate state after each reasoning step in the informal proof is explicitly presented in the Chain of State, and do not miss any step. If you find any step not presented in the Chain of State,  then output @@@RESULT@@@False@@@ and stop.

At the end, output all the informal states, and finally output @@@RESULT@@@True@@@ or @@@RESULT@@@False@@@ to indicate whether the Chain of State matches the informal proof as described and contains all intermediate states.

\end{tcolorbox}

\begin{table}[h]
  \centering
  \includegraphics[width=0.65\textwidth]{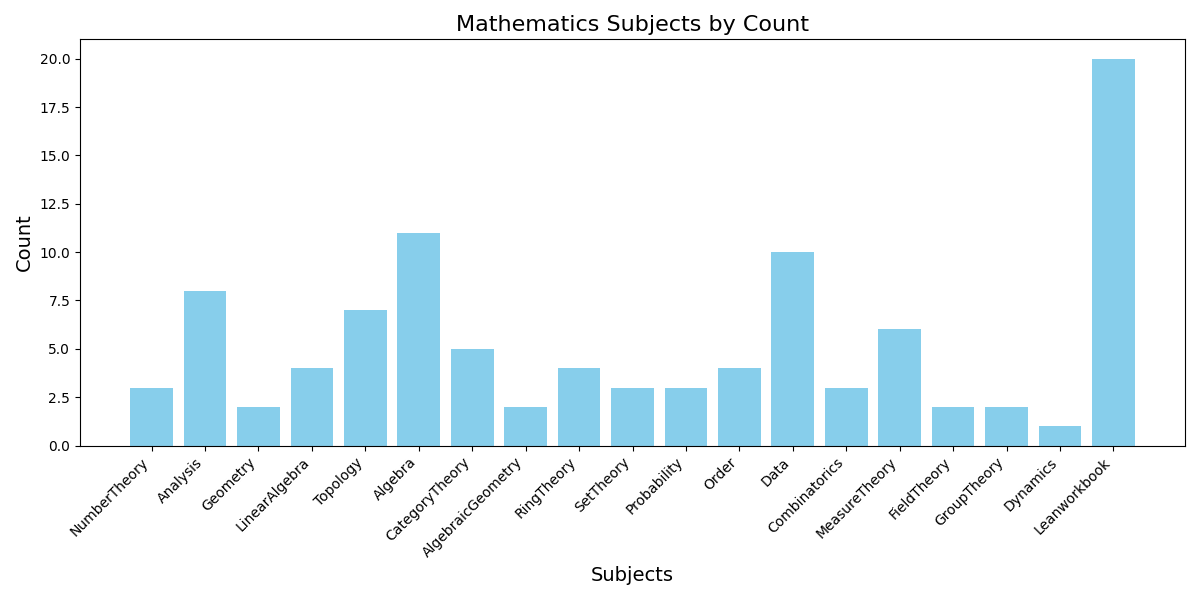}
  \caption{Composition of Mathlib problems}
  \label{fig:dist}
  \vspace{-8pt}
\end{table}

In addition to semantic verification, we apply a syntactic validation phase to ensure that all generated states conform to Lean’s elaboration and parsing rules. This step is carried out using \texttt{Leanjixia}, which invokes Lean’s elaborator on each state and flags malformed constructs or ill-typed expressions.

We perform CoS evaluation on three datasets: \texttt{MiniF2F-test}, \texttt{MiniF2F-valid}, and our constructed benchmark \texttt{CoS benchmark}. The CoS benchmark consists of 100 problems selected to cover diverse mathematical domains and proof styles. The dataset is constructed by extracting the formal CoS using \texttt{Leanjixia} from \texttt{mathlib} and \texttt{Leanworkbook}, followed by GPT-4o-based informal proof generation. Figure ~\ref{fig:dist} summarize the sources of the \texttt{CoS benchmark} across different datasets and its distribution over various domains in \texttt{mathlib}..

\subsection{Prompt for Chain of States Generator}
\label{Prompt_COS}
\begin{tcolorbox}[
  colback=gray!5,
  colframe=black!50,
  boxrule=0.4pt,
  arc=1mm,
  left=2mm, right=2mm, top=2mm, bottom=2mm
]
\small
Assume you are an expert in mathematics and Lean. Your task is to read the following formal statement, informal proof and detailed informal proof, and generate a formal chain of states for Lean. While generating the state list, you should insert explanations before each state where you think it's needed, clarifying how the next state is derived from the previous ones. In your output, the focus should be on the state list.

When generating a formal chain of states, you should adhere to the following principles:

\textbf{Principles of chain of states:}
Each state is a formal inference step in Lean. A state contains one or more goals. Each goal contains and only contains one symbol \(\vdash\), representing the inference. The name of each state should be ``State'' followed by a natural number starting from 0 and incrementing sequentially. Different states are separated by two newline characters. An example of a chain of states is:

\[
\begin{aligned}
\text{State 0:} & \quad a : \mathbb{R},\ b : \mathbb{R},\ c : \mathbb{R},\ d : \mathbb{R},\ e : \mathbb{R},\ h : a \leq b \ \vdash\ \operatorname{rexp}(a) \leq \operatorname{rexp}(b) \\
\text{State 1:} & \quad a : \mathbb{R},\ b : \mathbb{R},\ c : \mathbb{R},\ d : \mathbb{R},\ e : \mathbb{R},\ h : a \leq b \ \vdash\ a \leq b \\
\text{State 2:} &\quad \text{No Goals}
\end{aligned}
\]

You should refer to the detailed informal proof to generate the chain of states. Each state in your chain does not necessarily need to correspond one-to-one with each step in the detailed informal proof. If necessary, intermediate reasoning steps required for formalization should be added. For example, if the informal step says: 
$\text{``If } \frac{x + 9}{x - 1} = 1 \text{ and } x \ne 1, \text{ then } x + 9 = x + 1''$

then the chain of states should contain:

\[
\begin{aligned}
\text{State 0:} & \quad x : \mathbb{R},\ h_1 : \frac{x + 9}{x - 1} = 1,\ h_2 : x \ne 1 \ \vdash\ x - 1 \ne 0 \\
\text{State 1:} & \quad x : \mathbb{R},\ h_1 : \frac{x + 9}{x - 1} = 1,\ h_2 : x - 1 \ne 0 \ \vdash\ x + 9 = x + 1
\end{aligned}
\]

In each state of the chain, include basic assumptions about variables (e.g., \(x : \mathbb{R}\)) found in the formal statement. If the informal proof introduces new variables, you should also define them. Ensure that each transition from one state to the next reflects a simple and valid logical inference, and ultimately leads to the goal being proved.

\textbf{Principles of explanation:}
Insert short explanations before states where necessary. These should describe the goal in plain mathematical language and explain how it progresses. Not all states require an explanation, but all reasoning must be covered across the explanations.
\end{tcolorbox}

\section{DeepseekR1-based Adjacent States Prover}

To bridge adjacent formal states $\mathcal{S}_{\text{before}}$ and $\mathcal{S}_{\text{after}}$ with tactics, we develop a tactic generation pipeline based on \texttt{Deepseek-Reasoner}. Given the sequential nature of CoS, the model is prompted to synthesize minimal and efficient tactics capable of transforming the initial state into the final state. When a generation fails due to tactic errors, an automatic fallback mechanism is activated to diagnose the failure and generate revised tactics. Additionally, in cases where the tactic terminates at an incorrect intermediate state (i.e., a mismatch between actual and target final state), we introduce a state renewal mechanism to guide the model in completing the missing transformation.

\vspace{2mm}
\noindent
\textbf{Prompt for Primary Tactic Generation}
\begin{tcolorbox}[
  colback=gray!5,
  colframe=black!50,
  boxrule=0.4pt,
  arc=1mm,
  left=2mm, right=2mm, top=2mm, bottom=2mm
]
\small
You are an expert in Lean4 theorem proving. Your task is to analyze two given Lean4 proof states and generate the appropriate tactic(s) to transform from the initial state to the final state. Follow these requirements:

1. Carefully examine both the initial and final states, identifying the exact changes needed.  
2. Provide the minimal, most efficient tactic sequence that would accomplish this transformation.  
3. You can include any explanations, comments, or additional text outside the code block.  
4. If the transformation requires multiple steps, combine them with line breaks or use a single powerful tactic when possible.  
5. "No Goals" refers to the end of the proof.

\textbf{Example format for your response:}
\begin{lstlisting}[language=Lean]
tactic1; tactic2; tactic3
\end{lstlisting}
\end{tcolorbox}

\vspace{2mm}
\noindent
\textbf{Prompt for Error Recovery and Revised Tactic Generation}
\begin{tcolorbox}[
  colback=gray!5,
  colframe=black!50,
  boxrule=0.4pt,
  arc=1mm,
  left=2mm, right=2mm, top=2mm, bottom=2mm
]
\small
You are an expert in Lean4 theorem proving. Your task is to analyze two given Lean4 proof states and generate the appropriate tactic(s) to transform from the initial state to the final state, while considering any previously attempted tactics and their error messages.

1. Carefully examine both the initial and final states, identifying the exact changes needed.  
2. Review any previously attempted tactics and their error messages to avoid repeating mistakes.  
3. Provide the minimal, most efficient tactic sequence that would accomplish this transformation.  
4. If the previous attempt failed:
   - Analyze why the error occurred.  
   - Explain how the new tactic addresses the issue.  
   - Provide an improved tactic solution.  
5. Include any explanations, comments, or additional text outside the code block.  
6. If the transformation requires multiple steps, combine them with line breaks or use a single powerful tactic when possible.  
7. "No Goals" refers to the end of the proof.

\textbf{Please include the following in your output:}
- Previous failed attempt.  
- Error message received.  
- Analysis of why it failed.  
- Improved tactic solution.

\textbf{Example format:}
\begin{lstlisting}[language=Lean]
Previous failed attempt: `apply h`
Error: goal mismatch
Issue: `h` expects more assumptions
tactic1; tactic2; tactic3
\end{lstlisting}
\end{tcolorbox}

\vspace{2mm}
\noindent
\textbf{Prompt for State Renewal and Completion}
\begin{tcolorbox}[
  colback=gray!5,
  colframe=black!50,
  boxrule=0.4pt,
  arc=1mm,
  left=2mm, right=2mm, top=2mm, bottom=2mm
]
\small
I'm working on a Lean4 proof where we need to transform State A to State B. However, the current tactics only achieve State A to State C. Please analyze the proof context and:

- Carefully examine both the current state (State C) and target state (State B).  
- Identify what additional transformations are needed to bridge the gap between State C and State B.  
- Suggest the most appropriate tactics to complete the proof of State A.  
- Provide the complete tactic path from State A to State B.  

\textbf{Please output the final working tactic sequence inside triple backticks. Include brief explanatory comments for each tactical step where non-trivial reasoning is involved.}

\begin{lstlisting}[language=Lean]
tactic1; tactic2; ...
\end{lstlisting}
\end{tcolorbox}

\end{appendices}

\bibliographystyle{plain}
\bibliography{ref}

\end{document}